\documentclass[10pt]{article}
\usepackage{geometry}
\usepackage{xcolor}
\definecolor{graycolor}{gray}{0.9} 
\usepackage{microtype}
\usepackage{setspace} 
\usepackage[utf8]{inputenc}
\usepackage[english]{babel}
\usepackage{times}
\usepackage{array}
\usepackage{soul}
\usepackage{cite}
\usepackage[numbers,sort&compress]{natbib}
\usepackage{natbib}

\usepackage{titlesec} 
\titleformat {\section} [block] {\raggedright \fontsize{10}{10}\selectfont\bfseries} {\thesection. \space} {0pt} {}
\titlespacing {\section} {0pt} {12pt} {6pt}
\titleformat {\subsection} [block] {\raggedright \fontsize{10}{10}\selectfont\itshape} {\thesubsection .\space} {0pt} {}
\titlespacing {\subsection} {0pt} {12pt} {6pt}
\titleformat {\subsubsection} [block] {\raggedright \fontsize{10}{10}\selectfont} {\thesubsubsection .\space} {0pt} {}
\titlespacing {\subsubsection} {0pt} {12pt} {6pt}
\titleformat {\paragraph} [block] {\raggedright \fontsize{10}{10}\selectfont} {} {0pt} {}
\titlespacing {\paragraph} {0pt} {12pt} {6pt}

\usepackage{array} \newcommand{\PreserveBackslash}[1]{\let\temp=\\#1\let\\=\temp}
\newcolumntype{C}[1]{>{\PreserveBackslash\centering}m{#1}}
\newcolumntype{R}[1]{>{\PreserveBackslash\raggedleft}m{#1}}
\newcolumntype{L}[1]{>{\PreserveBackslash\raggedright}m{#1}}
\usepackage{lineno}
\usepackage{tabularx}
\usepackage{colortbl}
\usepackage{graphicx}
\usepackage{float}
\usepackage[export]{adjustbox}
\usepackage{caption}
\usepackage{fancyhdr} 
\usepackage{lastpage}
\usepackage{layout}
\usepackage{setspace} 
\usepackage{enumitem}
\usepackage{booktabs}
\usepackage{arydshln}
\usepackage{multirow}
\usepackage{color}
\usepackage{hyperref} 
\hypersetup{
	colorlinks=true,
	linkcolor=blue,
	filecolor=blue,
	urlcolor=black,
	citecolor=cyan,
}

\fancypagestyle{firstpage}{
    \setlength{\headsep}{2.2cm}
    
    \setlength{\footskip}{1.5cm}
    \fancyhf{}
    \lhead{\begin{table}[H]
        \centering
        \begin{tabular}{L{2.5cm}C{10cm}C{3.1cm}R{2cm}}
            \includegraphics[scale=0.035]{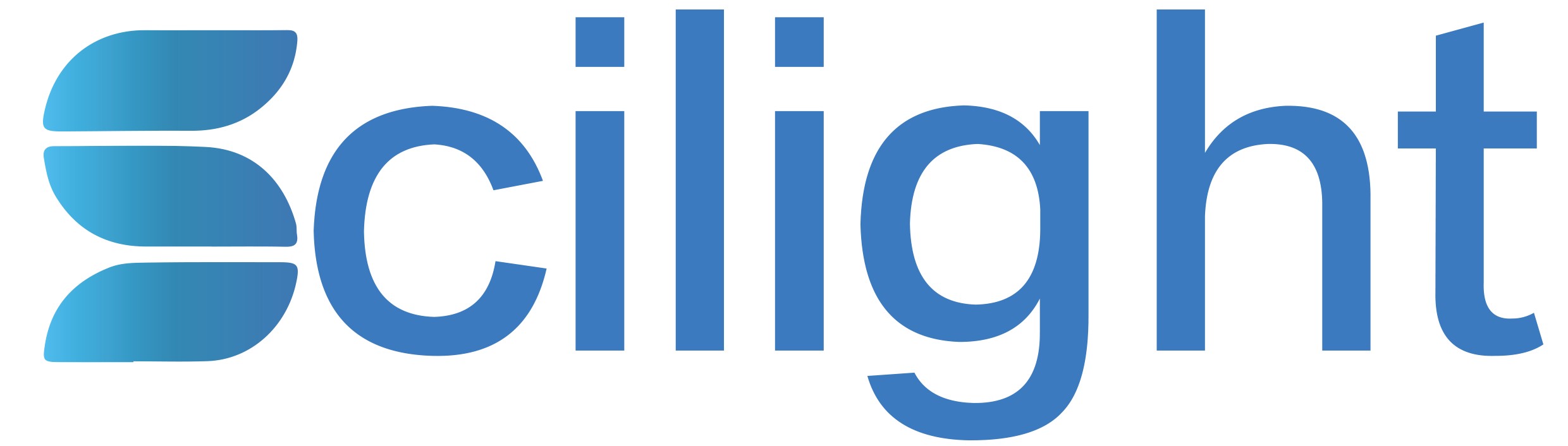} \vspace{-6pt}& \cellcolor{graycolor}\begin{tabular}[c]{@{}c@{}}\textit{Innovations in Space Research Technology}\\ \href{https://www.sciltp.com/journals/SpaceTech}{https://www.sciltp.com/journals/SpaceTech}\end{tabular} & \includegraphics[scale=0.029]{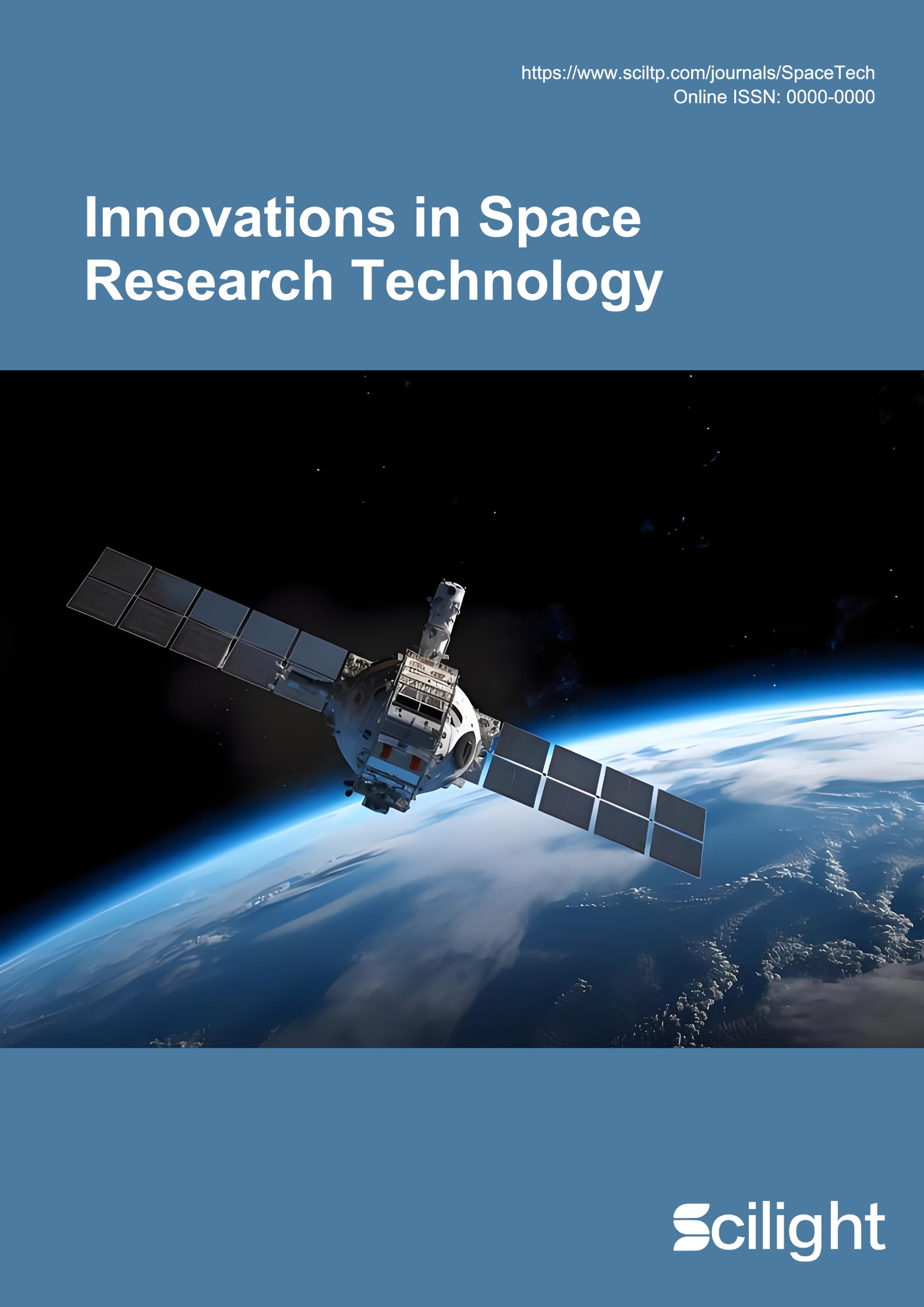} \vspace{-3pt}\\
        \end{tabular}
        \vspace{-22pt}
    \end{table}}
   
    \fancyfoot[C]{
        \vspace{-1.55cm}
        \begin{table}[H]
            \begin{minipage}[c]{0.15\columnwidth}
                \includegraphics[scale=0.5]{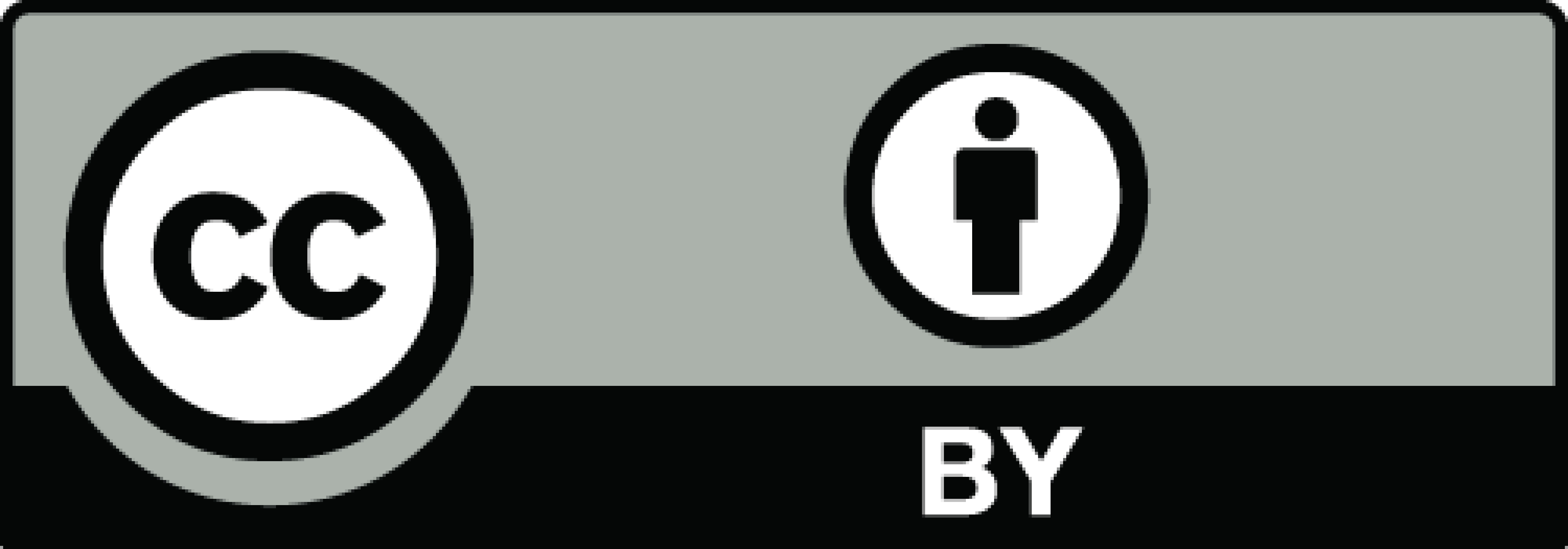} \vspace{1.1pt}
            \end{minipage}
            \hfill
            \begin{minipage}[c]{0.85\columnwidth}
                \scriptsize \textbf{Copyright:} © 2025 by the authors. This is an open access article under the terms and conditions of the Creative Commons Attribution (\mbox{CC BY}) license (\href{https://creativecommons.org/licenses/by/4.0/}{https://creativecommons.org/licenses/by/4.0/}). \\ \textbf{Publisher’s Note:} Scilight stays neutral with regard to jurisdictional claims in published maps and institutional affiliations.
            \end{minipage}
    \end{table}}
    \vspace{-0.55cm}
}

\begin{document}
\newgeometry{left=2.5cm, right=2.5cm, top=1.8cm, bottom=4cm}
	\thispagestyle{firstpage}
	\nolinenumbers
	{\noindent \textit{ Review}}
	\vspace{4pt} \\
	{\fontsize{18pt}{10pt}\textbf{Design and Scientific Prospects of the POLAR-2 Mission}  }
	\vspace{16pt} \\
	{\large Merlin Kole \textsuperscript{1,*}, Nicolas De Angelis \textsuperscript{2,\textdagger}, Jiang He \textsuperscript{3,\textdagger}, Hongbang Liu \textsuperscript{4,\textdagger}, Jianchao Sun \textsuperscript{3,\textdagger} Fei Xie \textsuperscript{4,\textdagger}, and Jimmy Zaid \textsuperscript{1,\textdagger }
	\vspace{6pt}
	 \begin{spacing}{0.9}
		{\noindent \small
			\textsuperscript{1}	\parbox[t]{0.98\linewidth}{University of New Hampshire, Space Science Center, Durham, 03824, New Hampshire, USA; merlinkole@gmail.com} \\
			\textsuperscript{2}	INAF Istituto di Astrofisica e Planetologia Spaziali, Via del Fosso del Cavaliere 100, 00133 Roma, Italy \\
            \textsuperscript{3}	State Key Laboratory of Particle Astrophysics, Institute of High Energy Physics, Chinese Academy of Sciences, Beijing 100049, China \\
            \textsuperscript{4}	Guangxi Key Laboratory for Relativistic Astrophysics, School of Physical Science and Technology, Guangxi University, Nanning 530004, China \\
		    {*}  \parbox[t]{0.98\linewidth}{Correspondence: merlinkole@gmail.com;} \\\\
			{\textdagger}  These authors contributed equally to this work. 			\vspace{6pt}\\
		\footnotesize	\textbf{How To Cite}: Kole, M. De Angelis, N., He, J. Liu, H. B. Sun, J. C. Xie, F. Zaid, J.; Design and Scientific Prospects of the POLAR-2 Mission. \emph{Innovations in Space Research Technology} \textbf{Year}, \emph{Volume}(Issue), Page Number. \href{https://doi.org/10.xxxx/xxx}{https://doi.org/10.xxxx/xxx}}.\\
	\end{spacing}

\begin{table}[H]
\noindent\rule[0.15\baselineskip]{\textwidth}{0.5pt} 
\begin{tabular}{lp{12cm}}  
 \small 
  \begin{tabular}[t]{@{}l@{}} 
  \footnotesize  Received: day month year \\
  \footnotesize  Revised: day month year \\
   \footnotesize Accepted: day month year \\
  \footnotesize  Published: day month year
  \end{tabular} &
  \textbf{Abstract:} The POLAR-2 mission consists of 3 instruments designed with the combined aim of producing a deeper understanding of Gamma-Ray Bursts. The mission will provide new insights regarding the geometries and emission mechanisms of the astrophysical jets which characterize these phenomena. To achieve this, POLAR-2 relies on polarization measurements and, for the first time will provide these using 2 separate polarimeter detectors. The first of these is a payload optimized to perform Compton polarimetry measurements in the $40-1000\,\mathrm{keV}$ energy range using a combination of plastic scintillators and silicon photo-multipliers. The development of this payload, the design of which is based on lessons learned from the POLAR mission, included optimization of plastic scintillator materials, their geometries and their wrapping. In addition, its development included detailed characterization, space qualification and radiation damage and mitigation strategies for the large number of silicon photo-multipliers included in the design. We will present these along with an overview of the readout electronics. These electronics were developed with flexibility in mind, as well as low cost and low power consumption. As such, its design is of interest beyond this polarimeter and is also used on the spectrometer instrument of POLAR-2 where it is used to read out an array of GAGG scintillators. This readout, in combination with a coded mask, allows this secondary instrument to provide detailed spectral measurements along with localization measurements of the observed gamma-ray bursts. The final instrument used in the mission aims to use gas-based detectors to perform polarization measurements in the keV energy region. The novelty of this design lies in its optimization for wide-field observations. In addition, it is specifically designed for transient source monitoring, capable of handling high fluxes, and its performance remains largely insensitive to rapid flux variations. The combination of the three instruments will allow to perform detailed spectral, localization and polarization measurements of these transient phenomena together for the first time. This paper will provide an overview of the technologies employed in the mission along with detailed predictions on its capabilities after its launch which is currently foreseen in 2027.   \\
\\
  & 
\end{tabular}
\noindent\rule[0.15\baselineskip]{\textwidth}{0.5pt} 
\end{table}

	\section{Gamma-Ray Polarimetry}

The field of high-energy astronomy employs photons with energies ranging from 100s of eV to 100s of GeV to study the Universe. Over the last 6 decades this has been made possible thanks to an array of space-based photon detectors, each one optimized to measure one or more of the 4 observables of a photon: their energy, incoming direction, arrival time or polarization. The first detectors, launched in the 60s and early 70s, primarily focused on spectral and localization measurements in the MeV energy range. This was soon followed by detectors capable of measuring the arrival time in detail, allowing, for example, the discovery of transient sources, such as Gamma-Ray Bursts (GRBs) by the Vela satellites \cite{1973ApJ...182L..85K}, as well as high energy emission from pulsars such as the Crab \cite{Bowyer1964, Tananbaum1971UHURU} . 

The discovery of high-energy emission from the Crab pulsar was followed within a decade by the first high-energy polarization measurements which focused on this object by OSO-8 \cite{OSO-8}. The OSO-8 mission was capable of measuring the polarization of the Crab pulsar in the soft X-ray energy range. These measurements showed a significant linear polarization of about 15 to $18\%$ at energies of 2.6 and 5.2~keV, with Polarization Angles (PAs) near 160 degrees. These angles were consistent with previous optical polarization measurements, which, in combination with the Polarization Degrees (PDs), strongly supported the hypothesis that the X-ray emission from the Crab Nebula is dominated by synchrotron radiation. 

These first results showed the potential for polarimetry to answer questions regarding astrophysical sources that cannot be addressed using spectrometry and timing measurements alone. In particular, polarimetry has shown to be a powerful tool to distinguish between various emission mechanisms and magnetic field environments. Yet, despite its early success, the next instruments capable of measuring polarization were not launched until the 1990s, while the first dedicated polarimeters were not launched until the 2010s \cite{overview_1}. An overview indicating some of the most important missions in the field of high-energy astronomy, with a particular focus on polarimeters, is shown in Figure~\ref{fig:hist}.

\begin{figure}
    \centering
    \includegraphics[width=0.8\textwidth]{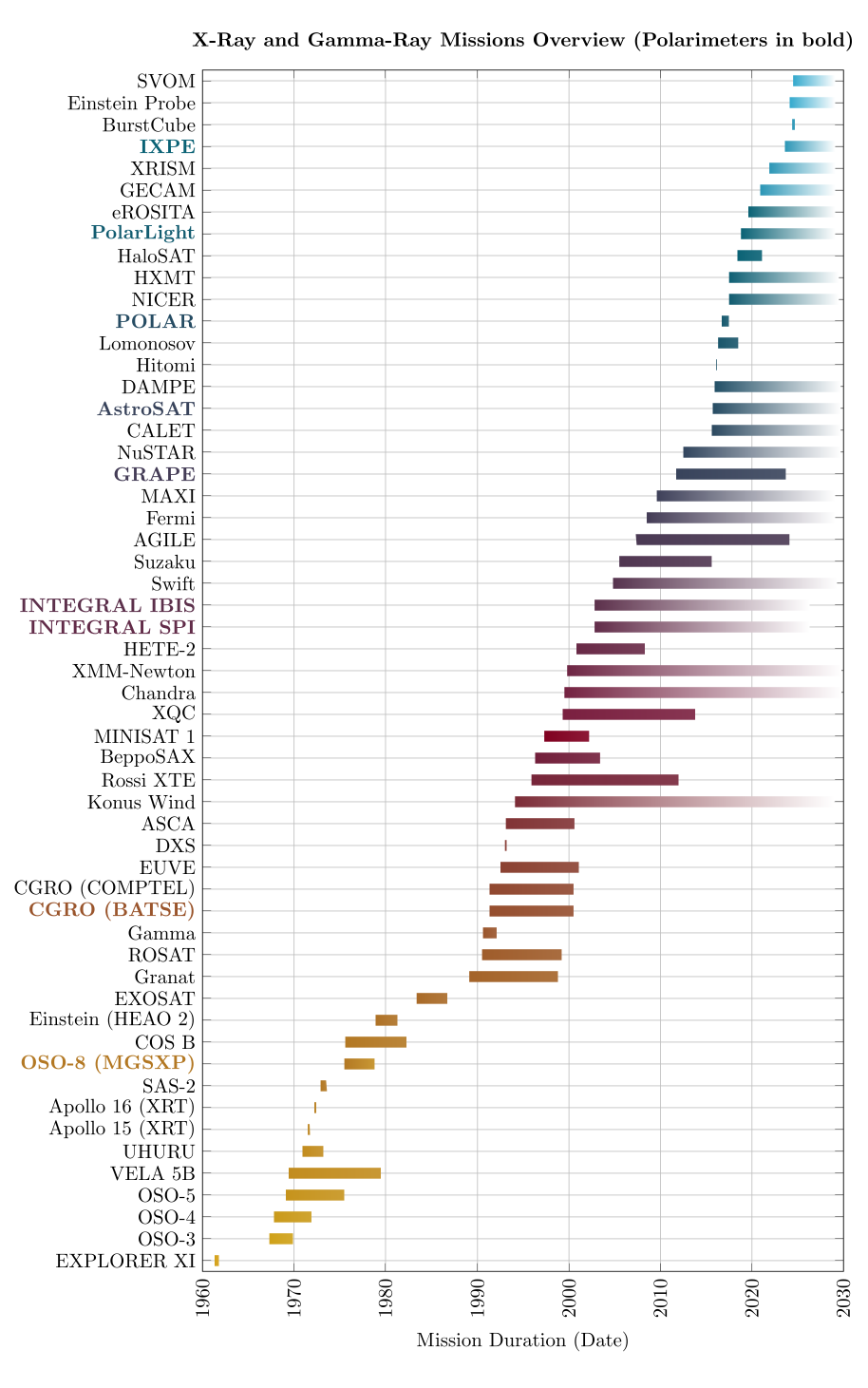}\hspace{3em}
    \caption{An overview of the history of high-energy astronomy with instruments capable of performing polarization measurements highlighted in bold. }
    \label{fig:hist}
\end{figure}

Like all instruments presented in Figure~\ref{fig:hist}, the polarimeters cover a wide energy range. The energy ranges covered by various instruments presented in Figure~\ref{fig:hist} are presented in Figure~\ref{fig:energy}. In the X-ray energy range\footnote{As no general definition of this exists, we here use the term X-rays for photons with energies below $\sim$20~keV, based on the different measurement technique required for polarimetry at such energies}, the OSO-8 mission was followed by Polarlight in 2019 \citep{Polarlight} and IXPE in 2021 \citep{10.1117/1.JATIS.8.2.026002}. Resulting in detailed polarization measurements of a range of sources such as the Crab, Vela PWN, and Cygnus X-1, to name a few \cite{Soffitta2024}. The field of X-ray polarimetry has, due to instrumental constraints discussed later in this paper, primarily focused on the measurement of steady-state continuous sources. This contrasts with the field of gamma-ray polarimetry where the primary focus has been on transient sources such as solar flares and GRBs. 

At gamma-ray energies, polarization measurements exist of only two non-transient sources: the Crab and Cygnus X-1. This is, as will be discussed in section \ref{subsec:polarization_measurement_concept}, a result of the low sensitivity of gamma-ray polarimeters which typically allows them to only observe the very brightest sources such as GRBs. In addition, whereas many sources are particularly bright at X-ray energies, GRBs typically show a peak in the emitted energy in the 100s of keV to MeV energy range \cite{Nava2011}. This, in combination with the wealth of information theorized to be contained in the polarization of the gamma-rays, has resulted in these phenomena being the primary target for the majority of the proposed gamma-ray polarimeters. 

Initial polarization measurements of GRBs were performed using non-dedicated instruments. This, as discussed in detail in \cite{MCCONNELL20171} and \cite{Gill2021}, resulted in an array of unreliable measurements. Polarization measurements in this energy range rely on the detection of asymmetries of the scattering angle of the photons. As such, anisotropies in the instrument sensitivity as well as chance coincident events can result in false polarization detections. Removal of instrument-induced effects can only be performed after detailed calibration of the detector, ideally with polarized beams prior to launch. As the initial instruments used for gamma-ray polarimetry were not designed for this purpose, no such calibrations existed and significant doubts on many of their measurements exist \cite{Gill2021}.

To mitigate these issues, dedicated GRB polarimeters were developed. The first of these was the Gamma-Ray Burst Polarimeter (GAP) mission \cite{Yonetoku2011_GAP}, which performed polarization measurements for 7 GRBs during about 2 years of operation \cite{Gill2021}. Although successful for a handful of GRBs, the GAP instrument did show the need for large sensitive areas to be capable of performing a significant number of GRB polarization measurements. This is due to the relatively low efficiency of such polarimeters. As will become more clear in section \ref{subsec:polarization_measurement_concept}, only a fraction of the detected photons can be used for polarization measurement purposes. As such, a large effective area is required compared to a spectrometer. This was in part mitigated by the POLAR mission \cite{PRODUIT2018259} launched in 2016. This mission, consisting of an array of 1600 plastic scintillators, was, through the use of an effective area approximately 10 times larger than GAP, able to perform 14 polarization measurements during its 6 months of operation \cite{Kole2020}. The results of POLAR provided some insight in the nature of GRBs, as will be discussed in more detail in section \ref{sec:science}. However, the relatively low PDs measured, as well as hints of time evolution of the polarization indicated that a yet higher sensitivity is required. In addition, the measurements indicated the importance of having detailed spectral and localization information of the GRB in order to reduce systematic errors.

The POLAR-2 mission is designed to not only be significantly more sensitive to polarization than its predecessors, it will also perform detailed measurements of the other 3 parameters of the incoming gamma-rays. It is designed to be capable of measuring the polarization of GRBs from several keV up to MeV energies. In this overview paper, we will first discuss the main scientific questions which POLAR-2 aims to answer in section~\ref{sec:science} along with the lessons learned from its predecessor POLAR. This is followed in section~\ref{sec:Design} by a detailed overview of the instrument designs as well as various lessons learned from the development which can be of use to the wider community. Finally, in section~\ref{sec:Prospects} we will present the scientific capabilities of the POLAR-2 mission and the impact it can have on the field. 

\begin{figure}
    \centering
    \includegraphics[width=0.8\textwidth]{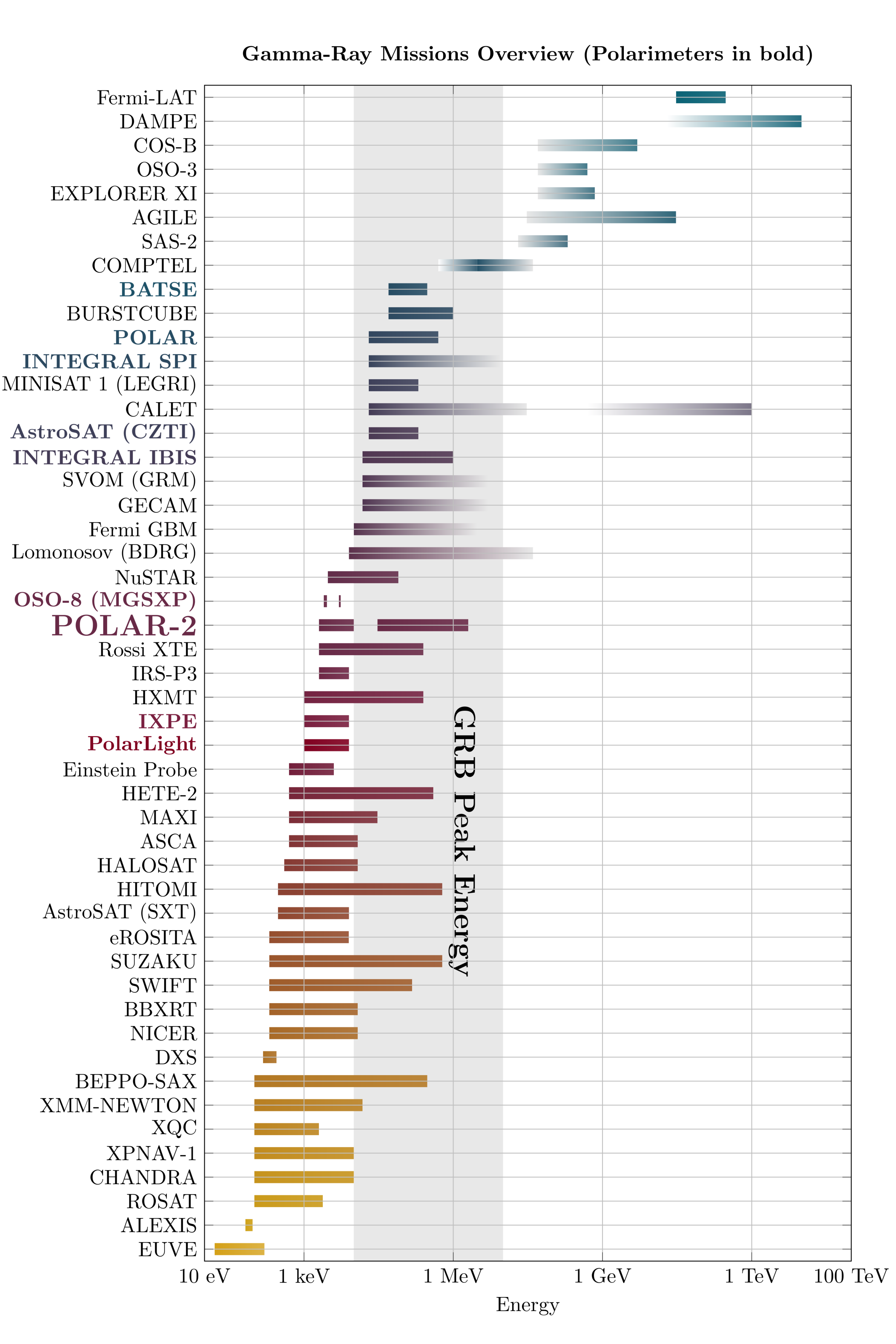}\hspace{3em}
    \caption{The energy ranges covered by the various instruments presented in Figure~\ref{fig:hist}, along with that of POLAR-2. The instruments with polarization capabilities are shown in bold, while the energy range in which the GRB spectrum typically peaks is shown in gray.}
    \label{fig:energy}
\end{figure}

\section{Scientific Motivation of the POLAR-2 mission}\label{sec:science}

GRBs are among the most energetic and electromagnetically brightest transient events in the Universe. They are characterized by a short, intense release of energy in the form of photons ranging from radio wavelengths to TeV energies. The most luminous phase of the burst is referred to as the ``prompt'' phase. This phase lasts from a fraction of a second to tens of seconds and shows a bimodal distribution in duration. As a result, GRBs are divided into 2 classes: short-hard (lasting $\sim 0.1$ s) and long-soft (lasting $\sim 30$ s) classes, related to distinct progenitors \cite{Kouveliotou+93}: compact object mergers in early type galaxies \cite{Gehrels+05, Barthelmy+05}(neutron star-neutron star or neutron star-black hole) and massive star collapses in star forming regions \cite{Woosley-93, Fruchter+06, Galama+98, Hjorth+03}, respectively. 

The connection of short GRBs to compact binary mergers was confirmed in 2017 thanks to the joint detection of a short GRB (GRB 170817A) with a gravitational wave event (GW 170817) \citep{GW170817_ligovirgo, Abbott_2017_counterpart}. The connection of longer GRBs with massive star collapses was initially made thanks to the discovery of afterglow emission at lower photon energies. The afterglow, which as the name suggests, succeeds the prompt emission, has been observed over the full electromagnetic spectrum ranging from radio waves \cite{Frail1997_GRB970508_radio} to TeV energies \cite{MAGIC2019,HESS2019}. It was the detection of the X-ray afterglow by BEPPO-SAX of GRB 970228 \citep{Costa1997_xray_afterglow}, which allowed for a precise localization which, in turn, allowed for the detection of the optical component of the afterglow \cite{vanParadijs1997_GRB970228_optical}. The joint observation localized the GRB to be within a distant galaxy (redshift of 0.695) in which a Type Ic supernova-like event was observed, thereby fully confirming the extragalactic nature of GRBs while also giving the first evidence connecting GRBs with supernovae. Over the last decades, this connection has been strengthened by a range of other measurements with the latest insights coming from X-ray prompt emission measurements by Einstein Probe. See for example \cite{Wu2025}.

Both in the case of long and short GRBs, the formation of a black hole is accompanied by the compact central engine which powers two ultra-relativistic jets \cite{Salafia2022}. It is within these jets that the prompt emission is theorized to be produced while the afterglow is a result of the jet material hitting the interstellar medium \cite{Paczynski1993}. The interaction of the jet with the material results in synchrotron emission which is initially bright at X-ray energies while, as the electrons lose energy, becomes softer.

While the afterglow emission is relatively well understood, the exact mechanisms responsible for energy dissipation and radiation in the prompt phase remain unresolved. Prompt GRB spectra are often described empirically by the Band function \cite{Band+93}, a smoothly broken power law which typically peaks in the range of 50~keV to 1~MeV \cite{Nava2011}. However, spectral modeling alone has failed to answer a range of important open questions regarding the nature of GRBs. The most important of which is probably: Which process is responsible for the gamma-rays which give these phenomena their name? Both synchrotron radiation and inverse-Compton scattering are theorized to be the prominent emission mechanisms. While for a detailed discussion on which questions polarimetry can answer we refer the reader to \citep{Gill2021}, the most important other open questions are:

\begin{itemize}
    \item What is the composition of GRB jets?
    \item What are the GRB energy dissipation and particle acceleration mechanisms?
    \item What is the magnetic field structure? 
\end{itemize} 

A range of different theoretical GRB models have been proposed over the last decades, each of which provides a different set of answers to the above questions. While all of these theoretical models can largely explain the observed spectral measurements of GRBs, their predictions on the observable polarization parameters vary significantly, thereby explaining the large interest in performing such measurements.

Regarding the question on the dominant emission mechanism, there are models in one class where synchrotron emission is the dominant mechanism \cite{Rees1992,Tavani1996}. On the other side, both in photospheric emission models \cite{Meszaros2000_photosphere} and Compton drag models \cite{Lazzati2000_ComptonDrag} the dominant mechanism is inverse-Compton scattering. This emission mechanism is, in theory, capable of producing $100\%$ polarized emission. However, this requires the photons to scatter with an angle of around $90^\circ$. In order to produce high levels of polarization the geometry of the last scattering interaction of the photons therefore needs to be very specific, with a large angle between the axis of the relativistic jet and the observer, to allow for high levels of polarization. 

\textbf{Photospheric emission models} attribute the GRB prompt gamma-rays primarily to thermal radiation released at the photosphere of the outflowing jet (the radius at which the jet becomes transparent to its own photons). The characteristics of the emission depend on the thermal properties and radiative transfer inside the jet photosphere, which, in combination with geometric and optical depth effects, lead to low levels of polarization. Photospheric emission models typically predict very low levels of polarization with a maximum of $\sim40\%$ predicted for GRBs observed far away from the jet axis \cite{Lundman_2018}. 

\textbf{Compton Drag} models also attribute the main emission to stem from inverse-Compton scattering, however, here these interactions are between the jet’s bulk relativistic electrons on external soft photons. Whereas in photospheric emission the photons have a thermal nature from the jet itself, here the photons have an intrinsically non-thermal nature. In addition, the emission is highly beamed and collimated, reflecting the relativistic jet interaction with an external photon field, rather than photons emerging from within the jet itself. The high beaming along with the non-thermal nature of the seed photons allows for higher levels of polarization \cite{2004MNRAS.347L...1L}. 

Within the category of \textbf{Synchrotron Models}, magnetic fields play a more prominent role within the jets than in the two previously mentioned models. The magnetic fields, which differ in their nature in the various sub-models, allow for the production of synchrotron emission from the electrons in the jet. Synchrotron emission is intrinsically polarized, while the overall PDs measured from these jets will depend on the magnetic field topography as well as the jet viewing angle \cite{2020MNRAS.491.3343G}. The highest levels of polarization are predicted by models with ordered magnetic fields, or \textbf{toroidal synchrotron models}. In such models, a toroidal magnetic field is ordered and wraps around the jet axis. The coherence of the field means that the synchrotron emission from relativistic electrons spiraling in these fields will have a well-defined preferred polarization direction over the visible jet surface. The ordered magnetic field also implies that the composition of the jet is Poynting flux dominated, whereas the jets in all the other major models are matter dominated.

Synchrotron emission is also expected from \textbf{random synchrotron models} where the magnetic fields are produced within the shocks \cite{Granot_2003}. The energy dissipation in the jet therefore occurs through internal shocks, while for ordered magnetic fields it is mainly in the form of magnetic reconnection. Different types of random synchrotron models exist which differ regarding the geometry of the shock-generated magnetic fields. The two most common variants are the "parallel" and "perpendicular" field models. In the parallel case, the magnetic field is randomly oriented in all three spatial dimensions (3D isotropic random), as might occur ahead of, or far behind, a relativistic shock where turbulence is fully developed. Since the magnetic fields within the observer's field of view are randomly distributed, the polarization angles from different patches can cancel out, thereby producing a typically low net polarization. 

In the perpendicular case, the magnetic field is randomly oriented but confined to the plane perpendicular to the jet axis. Although random in orientation in the local shock plane, the structure breaks perfect isotropy along the jet axis. Due to relativistic beaming, the net polarization does not completely cancel. The projected magnetic field onto the plane of the sky is organized enough over the visible region that a modest net polarization survives. Finally, the two random synchrotron models do not necessarily exclude each other, meaning one can have models where both types of magnetic field structures exist, thereby producing low to moderate polarization angles.

\begin{figure}[h]
    \centering
    \includegraphics[height=.40\textwidth]{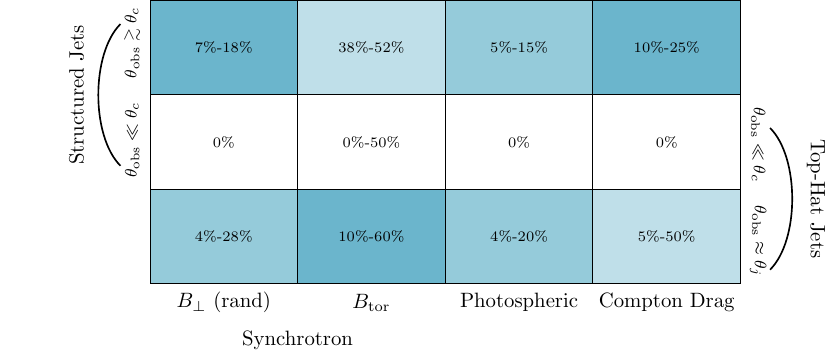}
    \caption{The predicted values of the PD for different theoretical models. The Compton Drag, Photospheric model, as well as two synchrotron models are shown. The PD is shown both for on-axis observations as well as for observations near the jet edge in case of a Top-Hat jet structure and a structured jet structure. The figure based on that from \cite{Gill2021}.}
    \label{fig:Models}
\end{figure}

An overview of the PDs predicted by these models is shown in Figure~\ref{fig:Models} for 3 different geometries. First there is the case a small viewing angle compared to the jet opening angle. The other two geometries are for a larger observation angle where the jet structure differs for the two geometries. In one case (the bottom row) the jet has a top-hat structure, meaning it has a near uniform Lorentz factor along the jet radius which drops abruptly when going beyond the jet radius. In the case of a structured jet model (top row), the Lorentz factor instead drops more smoothly along the jet radius.

To date over 50 GRB gamma-ray polarization measurements have been published, a detailed discussion of these is presented in \cite{Gill2021}. Although the earliest prompt polarization measurements, such as those by RHESSI \cite{Coburn_Boggs_2003, Rutledge2004, Wigger2004}, found high levels of polarization, these were often the result of inaccuracies in the analysis or due to instrumental effects \cite{MCCONNELL20171}. With measurements by dedicated instruments like POLAR and GAP, as well as instruments calibrated for polarization measurements prior to launch such as AstroSAT CZTI \cite{2015A&A...578A..73V}, a more consistent picture started to emerge. The GAP instrument \cite{GAP} reported a total of 7 GRB polarization measurements. Out of these 2 were consistent with a relatively high PD, while the others were consistent with a low or unpolarized emission \cite{Gill2021}. The 14 GRB measurements by POLAR \cite{Kole2020} are also all consistent with a low or unpolarized emission, with some strong upper limits set to PD levels around $40\%$. Early reports by the AstroSAT CZTI collaboration instead found a high PD for the majority of the GRBs measured \cite{CZTI_nonsense}. Later, through updated analysis, the results were found to be more consistent with those from POLAR \cite{Chattopadhyay_2022}. Although the majority of the measurements were no longer significantly constrained after this analysis update.

Overall, the image of time and energy integrated polarization measurements appears to favor PD levels below $\sim40\%$, although some outliers exist. It should be noted that, based on the discussion above, all of the models discussed allow for a range of PD values. In order to distinguish between them, it is therefore important to not only measure the PD with a higher precision than what has currently been achieved, but also to perform such measurements for a large number of GRBs. In order to advance the field a more sensitive detector is therefore required.

\subsection{Time Resolved Studies}

The majority of GRB emission models include jets that are symmetric around their central axis. This symmetry results in the clear prediction that during the GRB the PA of the emission can either flip by $90^\circ$ or not at all \cite{2020ApJ...892..141L, Gill2021_a}. Measurements of the time evolution can therefore be used to directly confirm or disprove such a symmetry. When assuming a top-hat jet and symmetry around the axis, the three magnetic field configurations, namely $B_\perp$ , $B_\parallel$ , and $B_{tor}$ predict a distinct polarization evolution over an emission pulse in the prompt emission. For an on-beam observer both $B_\perp$ and $B_\parallel$ would show $90^\circ$ changes in the PA while the $B_{tor}$ field case either shows two such changes over the pulse duration or none depending on outflow and geometric configurations \cite{Gill2021_a}. When the received emission is off-beam, only the $B_{tor}$ shows $90^\circ$ PA changes whereas both $B_\perp$ and $B_\parallel$ fields show a steady PA. Photospheric emission models show no polarization when viewed on-axis, while when viewed off-axis it allows only a flip in PA by $90^\circ$ \cite{Lundman_2018}. If such a change in the PA would occur for any such models, the resulting PD for a time-integrated analysis would be lowered as the changing PA smears out the polarization signal. The typically low PD values found in the time-integrated analysis by POLAR and AstroSAT thereby could simply be an artifact of a changing PA during the GRB.

To date, only a limited number of GRB polarization measurements exist where time-resolved studies were performed. The reason for this is simply the lack of sensitivity. When dividing the already statistically limited data into time bins it further decreases. One of the first time resolved study was performed by GAP for GRB 100826A where, by dividing the long GRB into 2 time bins, a change of the PA consistent with $90^\circ$ was found with a PD of around $30\%$. This while a time integrated analysis did not show any significant polarization levels \cite{GAP}.  

Studies using the POLAR data, such as those in \cite{Burgess2019, Kole2020}, showed hints of time evolution of the polarization angle for two GRBs: 170114A and 170101A. For both GRBs the data was most consistent with a PD of $\sim30-40\%$ with a rapidly changing polarization angle. It is worth noting that both these GRBs were single-pulse GRBs, while no significant evolution was found for any of the other GRBs in the POLAR catalog. However, even for the two single-pulse GRBs the results were not constraining enough to draw strong conclusions. Finally, several time-resolved studies using the AstroSAT CZTI data were performed, where again no strong conclusions can be drawn \cite{time_nonesense} as the PD is compatible with an unpolarized flux for most of the studied time bins. The lack of significant time resolved polarization measurements further indicates the need for more sensitive polarimeters.

\subsection{Energy Resolved Studies}

The PD can naturally be expected to be energy dependent. A clear example of this is synchrotron emission where the polarization degree, as well as the emitted energy spectrum will depend on the energy of the radiating electrons. For synchrotron emission models on can therefore expect the polarization degree to vary along with the evolution of the spectral parameters. As such effects are relatively small, and require a high amount of statistics, such measurements would only be possible for extremely bright GRBs.

A more probable application of energy-resolved polarimetry is to identify whether different emission mechanisms dominate at different energies. This is of particular interest for GRBs which show a clear thermal peak in the energy spectrum, such GRBs make up about $15-20\%$ of GRBs. The thermal peak is generally theorized to be produced through Comptonization of low-energy photons at distances where the jet has become transparent \cite{Lundman_2018}. It is therefore, generally expected to be lowly polarized. However, if the emission outside of this peak region has a synchrotron origin, one would expect a strong energy dependence on the polarization. Predictions by \cite{Lundman_2018} indicate that non-Comptonized synchrotron emission can be observed at lower energies which therefore remains polarized, especially in the X-ray region (below $\sim20\,\mathrm{keV}$), while above $\sim100\,\mathrm{keV}$ the emission is fully unpolarized. Verifying such models would therefore require polarization measurements that cover both the X-ray and gamma-ray energy ranges. 

To date, no significant energy-resolved GRB polarization studies have been performed. A framework for the analysis was set up in \cite{NDA_thesis, energy_resolved_icrc}. This work properly took into account the energy dispersion of the polarimeter using forward-folding, which is required as the polarization is dependent on the photon energy, not the measured energy. As, similar to time resolved analysis, the statistical limitations further increase when dividing the data in energy bins no significant conclusions could be drawn from this work. Similarly, for polarization measurements from other instruments the statistics are currently too limited to allow for energy-dependent studies to draw strong conclusions. These studies therefore show the need for significantly more sensitive detectors to be developed. Additionally, as the energy dependence occurs at 10's of keV, it is important to have an instrument which, unlike POLAR, has significant polarization sensitivity at such energies. A more sensitive instrument with a larger dynamic range is therefore required.

\subsection{Multi-Messenger Studies}

As is clear from Figure~\ref{fig:Models}, the expected polarization degrees depend not only on the underlying emission model, but also on the viewing angle. Having a measurement of this angle, which can be achieved through afterglow measurements \cite{Rossi}, therefore significantly increases the discriminating power of polarization measurements. It is therefore important to maximize the probability of having afterglow measurements of the GRB after its prompt emission has been detected by POLAR-2. To achieve this, an instrument like POLAR-2 should ideally be equipped with a real-time alert system which allows to send transient alerts to the GRB community as quickly as possible. Secondly, the afterglow observations at, for example, optical wavelengths, require a detailed position on the sky to be initiated. It is for this purpose that the POLAR-2 mission will contain not only polarimeters, but also the Broadband Spectrometer Detector (BSD) which will be discussed in more detail in section \ref{sec:BSD}. This instrument will, through the use of a coded mask, be capable of providing sub-degree measurements for a large number of the GRBs for which the polarimeters will perform polarization measurements. This, in addition to being foreseen to be equipped with a real-time alert system, allows to maximize the probability for the GRB polarization measurements to be accompanied by observations capable of constraining the jet geometry.

Finally, joint observation data from gravitational wave detectors has been identified as a promising source of data to further increase the potential of GRB polarization data \cite{Kole_GW}. This study indicates that polarization measurements of the gravitational wave allow to constrain the 2 viewing angles with which the progenitor binary pair of short GRBs are observed. As discussed in detail in \cite{Kole_GW}, these angles are directly connected to the allowable PA of the GRB polarization. While for POLAR-2 a handful of joint observations with LIGO are expected during the O5 run, the study shows that the current generation of gravitational wave detectors is not sensitive enough to constrain the observational angles of the GRB source. However, the tilting angle ($\iota$) of the binary pair can be constrained and, in most models, is strongly correlated with the PD of the GRB polarization.

\section{Instrument Design}\label{sec:Design}

\subsection{Polarization Measurement Concept}\label{subsec:polarization_measurement_concept}

The measurement of the polarization of X-ray or $\gamma$-ray photons is an indirect one. For spectral measurements the energy of the incoming photon is typically related through a near-linear equation to the energy deposited in the detector. In high-energy polarimetry, the direction of the polarization vector of an incoming photon is related to the probability distribution of the emission direction of the photo-electron in photo-absorption, or its azimuthal scattering direction when employing Compton scattering. At even higher energies, it is instead related to the production plane of the electron-positron pair. The measurement of the PD and PA for an incoming photon flux therefore requires the measurement of a distribution of these parameters, a process which is generally statistically hungry and prone to instrument-induced biases \cite{MCCONNELL20171}.

\begin{figure}[h]
    \centering
    \includegraphics[height=.40\textwidth]{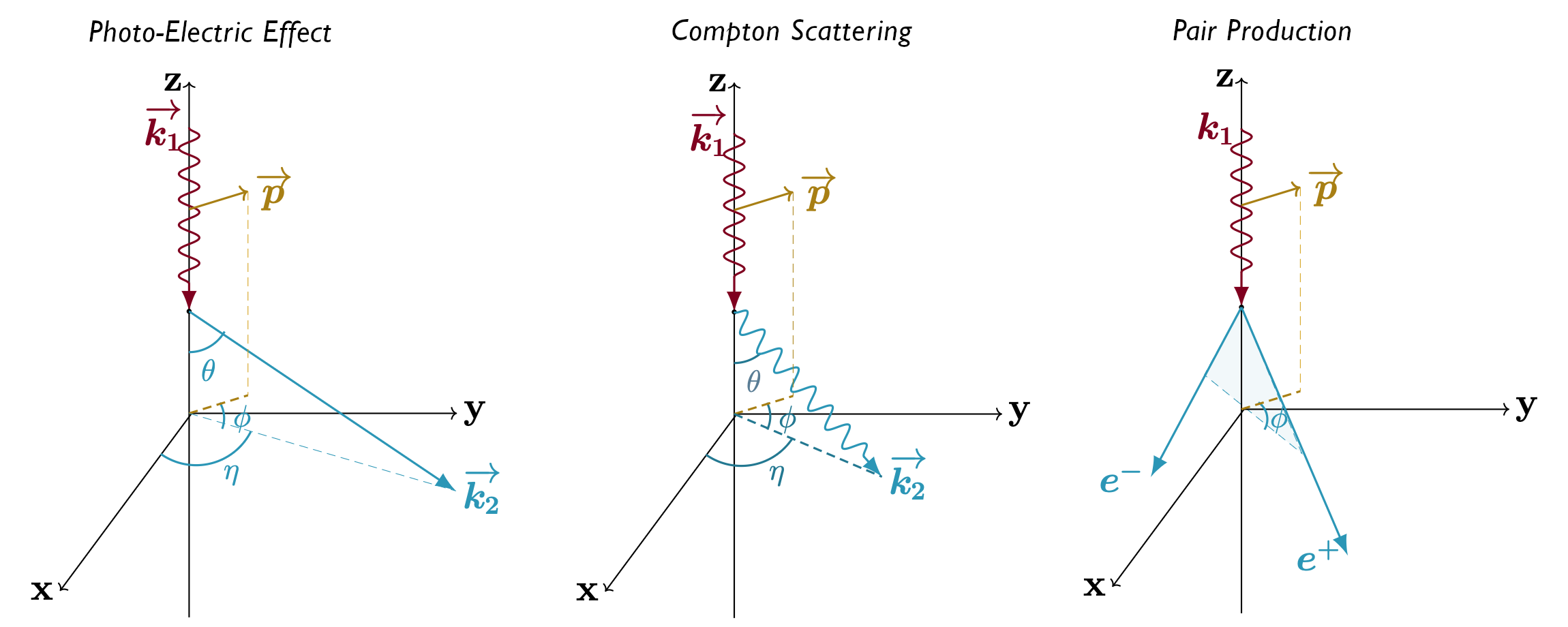}\hspace*{0.1cm}
    \caption{Illustration of the dependence of the polarization vector on the secondary particles for the photo-electric effect (left), Compton scattering (middle) and pair production (right). Adapted from \cite{NDA_thesis}, with permission.}
    \label{fig:Interaction}
\end{figure}

The interaction principles for the three processes are shown in Figure~\ref{fig:Interaction}. In this work, we will solely focus on the photo-absorption process, which is employed in the Low-energy Polarimetry Detector (LPD) of POLAR-2, and Compton scattering which is employed in the High-energy Polarimetry Detector (HPD). In the former it is the azimuthal direction of the outgoing electron which shows a dependency on the polarization vector of the incoming photon $\vec{p}$. In the differential cross section for this interaction mechanism a dependency exist on $\phi$ which is defined as the 
azimuthal angle between the polarization vector of the incoming photon $\vec{p}$, see Figure~\ref{fig:Interaction}, and the projection of the velocity vector of the secondary electron $\vec{\beta} = \vec{v}/c$ within the plane of the  
momentum vector $\vec k_1$ of the photon,
\begin{equation}\label{eq:phi_dsigma_dOmega}
    \frac{d\sigma}{d\Omega}\propto \cos^2\phi\, \quad,\quad 
    \phi  =\cos^{-1}\left(\frac{\vec{k}_2\cdot\vec{p}}{k_2 p\sin\theta}\right)\ ,
\end{equation}

Here $d\Omega=\sin\theta d\theta d\phi$ is the unit solid angle and the polar angle $\theta$ is given by 
$\cos\theta=\hat{\beta}\cdot\hat{k}_1=\hat{k}_2\cdot\hat{k}_1$. 

The result of this dependency is that, when measuring the outgoing momentum vector of the secondary electron, it will show a modulation with a period of $180^\circ$ where the amplitude is directly related to PD, while the maximum is related to the PA of the incoming photon flux.

Measuring the polarization by measuring the distribution of the outgoing momentum vector of the secondary electron in photo-absorption is complex, but can be performed using gas-based detectors in the $1-10\,\mathrm{keV}$ energy range. It is therefore employed in the LPD of POLAR-2. At higher energies the Compton scattering cross section becomes dominant and, therefore polarization measurements using this process become more efficient. For this process a dependency of the differential cross section exists on $\phi$, defined here as the angle between the polarization vector of the incoming photon $\vec{p}$ and the projection of the momentum vector 
of the outgoing photon $\vec{k}_2$ on the plane of the polarization vector $\vec{p}$. Here 
$\phi = \cos^{-1}\left(\frac{\vec{k}_2\cdot\vec{p}}{k_2 p\sin\theta}\right)$ as in Eq.~(\ref{eq:phi_dsigma_dOmega}), is
\begin{equation}\label{eq:KN}
    \frac{d\sigma}{d\Omega} = \frac{r_o^2}{2}\frac{E'^2}{E^2}\left(\frac{E'}{E}+\frac{E}{E'}-2\sin^2\theta \cos^2\phi\right).
\end{equation}
Here $r_0 = e^2/m_ec^2$ is the classical electron radius with $e$ being the elementary charge, $E$ is the initial photon energy, $E'$ the 
final photon energy, $\theta=\cos^{-1}(\hat{k}_2\cdot\hat{k}_1)$ the polar scattering angle.

The result of this dependency is similar to that for the photo-electric effect: A measurement of the angle $\phi$ for an incoming photon flux will result in a harmonic distribution with a period of $180^\circ$ with the amplitude being dependent on the PD and the phase on the PA. 

It is however important to note that the angle $\phi$ cannot be measured directly as it requires knowledge on the direction of $\vec{p}$. Therefore the angle typically measured is that between $\vec{k_2}$ and a known axis defined in the instrument. Here, as illustrated in Figure \ref{fig:Interaction}, this is the angle $\eta$ measured with respect to the x-axis. This results in a phase shift of the measured distribution. The same principle holds for photo-absorption measurements. An example of the types of distributions which would be measured for a polarized and an unpolarized flux are shown in Figure~\ref{fig:mod_examples}.

\begin{figure}[h]
    \centering
    \includegraphics[height=.35\textwidth]{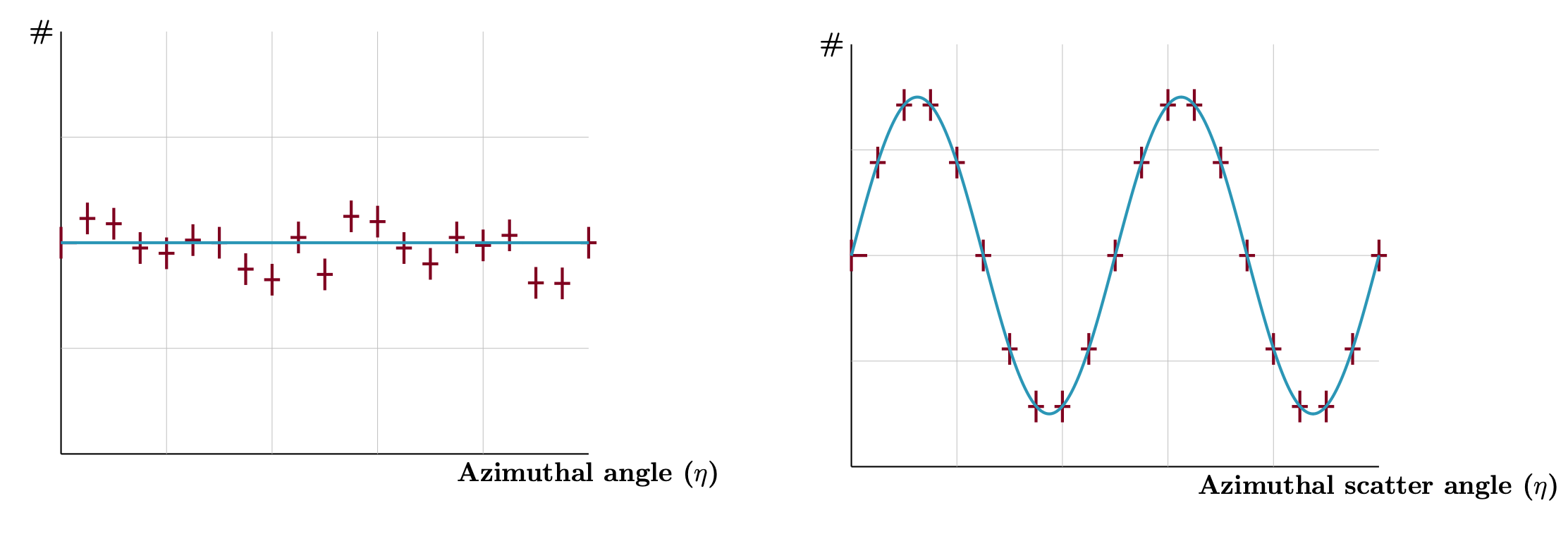}\hspace*{0.1cm}
    \caption{An illustration of the measured distributions of the electron emission angle (for the photo-electric effect), or scattering angle (for Compton scattering) for an unpolarized flux (left) and a polarized flux (right).}
    \label{fig:mod_examples}
\end{figure}

As can be derived from equation \ref{eq:KN} the relative amplitude of the harmonic depends on the energy of the incoming photon. In addition, a dependency will exist on the sensitivity and efficiency of the detector. Knowledge of the relative amplitude one can achieve for a PD of $100\%$, referred to as $M_{100}$ is required to retrieve the PD through $PD=\frac{M}{M_{100}}$ where M is the relative amplitude for a measurement. The value of $M_{100}$ can be measured for discrete energies for a detector prior to launch, however, producing it for the full energy range of the detector requires the use of detailed Monte Carlo simulations. An error on the simulated $M_{100}$ from such simulations will therefore linearly correlate to an error on the measurement of PD. This indicates the importance of well calibrated instruments where detailed calibration results are used to validate the instrument simulations in as much detail as possible \cite{Gill2021}. 

The issues stemming from the reliance on simulations become further clear when realizing that the value of $M_{100}$ also varies with the incoming direction of the photon beam with respect to the instrument. As the POLAR-2 instrument is designed for GRBs, of which the incoming direction is not known prior to the measurement, this dependency needs to be understood as well in detail. Thereby producing the need for off-angle calibrations of the polarization response as well. It should be noted here that the off-axis response of polarimeters is highly complex and requires both a detailed understanding of the instrument, as well as the physics, detailed studies of the dependencies can be found in \cite{Fabio} and \cite{Zuke,Zaid}.

One figure of merit which is often used to define the sensitivity of a polarimeter to measure a source is the Minimal Detectable Polarization (MDP). The definition of the MDP is as follows \cite{weisskopf_x-ray_polarization}:

\begin{equation}
    \mathrm{MDP_{CL}} = \frac{\sqrt{-2\mathrm{\ln(1-CL)}}}{M_{100} R_S} \sqrt{\frac{R_S + R_B}{T}}
\end{equation}

The MDP here represents the value of the incoming PD which can be distinguished in an observation from an unpolarized flux with a given confidence level (CL).  In this $M_{100}$ is the modulation factor as measured for PD=$100\%$ as used before. $R_S$ is the signal rate detected from the source, $R_B$ the background rate and  $T$ the observation time. 

For the purpose of GRB measurements the $T$ is defined by the length of the event. $M_{100}$ is instead a parameter which depends both on the energy and incoming angle of the GRB and should be optimized in the instrument design. Finally, the $R_S$ depends on the effective area of the instrument and $R_B$ is typically related to the geometric volume of the detector. In order to allow for sensitive measurements the polarimeter should be optimized to have a large $M_{100}$ as the MDP scales linearly with this value. The value of $M_{100}$ depends strongly on the precision with which the angle $\eta$ can be measured. This value can therefore be optimized by finely segmenting the readout area of the polarimeter, thereby improving the angular resolution. In addition, a large effective area is required to maximize the number of photons detected from these bright but, relatively short events. Optimization of the effective area can be achieved simply by producing a large geometric area while optimizing the probability to measure the required interaction mechanism. For a Compton polarimeter, the latter can be achieved by selecting detector materials with a low Z, which increases the probability for photons to Compton scatter rather than undergo photo-absorption. As for the purpose of POLAR-2 the scientific objects of interest are transient events, it is furthermore important to note that the effective area should not just be large for on-axis observations but should remain significant for off-axis measurements as well to maximize the number of GRBs for which polarization measurements can be performed.

\subsection{HPD Design}

The POLAR-2 mission will consist of two payloads, the HPD that will be introduced in this section and the Spectrocopy and Polarimetry Detector (SPD) which will be discussed in section \ref{sec:BSD}. Both of these will be placed on the China Space Station (CSS). This has the advantage that they can rely on a relatively large amount of power produced by the space station, compared to for example a free-flying satellite, while the station also handles the data transfer for the mission. As a result, the two payloads can be relatively large (exceeding 100 kg), have power consumptions of up to 300 W and can produce tens of GB of data per day.

The HPD is a single-phase Compton polarimeter whose design is based on the legacy of the POLAR mission \cite{PRODUIT2018259}. It consists of a segmented array of scintillators that exploits the correlation between the polarization vector of a photon and its azimuthal Compton scattering direction to determine the polarization parameters (degree and angle) of a given source in a statistical manner, as explained in section~\ref{subsec:polarization_measurement_concept}. As in the case of POLAR, it is based on a modular design, with a single polarimeter module consisting of an 8$\times$8 array of elongated plastic scintillators read out by a multi-channel light sensor. While POLAR was made of 25 modules arranged in a square array, POLAR-2 will be composed of 10$\times$10 polarimeter modules, increasing the number of channels from 1600 to 6400. In addition to increasing the number of modules by a factor of 4, the polarimeter module design itself was improved in order to further enhance the HPD sensitivity compared to POLAR. A detailed model of the POLAR-2 polarimeter module design is shown in Figure~\ref{fig:hpd_module_design}, while the overall instrument design is depicted in Figure~\ref{fig:hpd_design}. We describe hereafter the major design changes that were brought to the polarimeter module to strengthen POLAR-2's scientific abilities.

\subsubsection{Scintillator Design}

Firstly, the plastic scintillator bars were shortened from 176~mm to 125~mm to improve the signal-to-noise ratio (SNR) of the polarimeter, finding a good compromise between SNR and overall sensitivity \cite{SNR_TN_POLAR-2}. Indeed, while the background scales with the total volume of the scintillators, the effective area of the instrument goes with the surface in the plane orthogonal to the source's direction. Shortening the bars therefore allows for increasing the SNR while also slightly improving the sensitivity to polarization (increasing the $M_{100}$)  by reducing the range of possible polar scattering angle to be centered closer to the region for which the dependence of the polarization on the azimuthal scattering angle distribution is largest (near $90^\circ$). The scintillators were also made wider, from 5.8$\times$5.8~mm$^2$ to 5.9$\times$5.9~mm$^2$, thanks to a new wrapping technique, designed to be optimal for the POLAR-2 HPD, allowing for individual wrapping of the bars while also reducing both dead spaces between channels and optical crosstalk \cite{POLAR-2_optical}. While the POLAR bars were truncated from 5.8$\times$5.8~mm$^2$ down to a surface of 5.0$\times$5.0~mm$^2$ on both extremities for mechanical reasons, the POLAR-2 scintillators are kept straight up to the extremities, increasing the contact surface with the sensors by about 40~\%, thereby improving light collection efficiency. For POLAR, the truncation also provided a mechanism to align the scintillators with the Multi-Anode PhotoMultiplier Tubes (MAPMT) channels which read them out, within a 50~$\mu$m precision. For POLAR-2, a similar alignment is achieved but now using a highly precise 3D-printed scintillator alignment grid with 200~$\mu$m-thick separations between bars \cite{POLAR-2_optical}.

\subsubsection{Optical Sensors}

Significant progress has been made in the field of light sensing for particle physics and astronomical applications in the last decades. This comprises the maturing of Silicon PhotoMultipliers (SiPMs) that became widely used in ground-based high-energy physics experiments and are starting to be used in space-based missions as well. Compared to photo-multiplier tubes (PMTs), SiPMs offer a better mechanical robustness, a more compact design, and necessitate much lower bias voltages. These advantages make them very promising for space applications over PMTs because of the constraints induced by the harsh launch conditions and limited mass and volume constraints. The plastic scintillator readout sensor was therefore upgraded from the MAPMTs used in POLAR, to SiPM arrays for POLAR-2. In addition to the above-mentioned advantages, SiPMs typically offer a higher light sensitivity than PMTs, leading to a photo-detection efficiency twice that of the POLAR PMT's quantum efficiency.

This upgrade allowed great improvement in light collection efficiency at the output of the scintillator, which is crucial to improve the instrument sensitivity at low energy, since softer X-rays will produce a small amount of optical light in the scintillators. This, combined with optimizations based on thorough optical characterization and simulation of the polarimeter module design \cite{POLAR-2_optical}, allowed to increase the amount of detected optical light per unit of deposited energy by over a factor of 5, from 0.3~photoelectrons (p.e.)/keV to 1.6~p.e./keV. This great enhancement of light yield has a significant implication on the instrument's sensitivity, especially at low energies where the sources are the brightest, as discussed in section~\ref{sec:Prospects}. Another benefit of using SiPMs over MAPMTs is the much thinner entrance protective layer, from 1.5~mm for POLAR MAPMT's borosilicate glass window to 0.1~mm for the epoxy resin of POLAR-2's SiPMs. Together with a thinner optical coupling pad between the sensor and the scintillators block, whose thickness was reduced from 0.7~mm to 150~$\mu$m by directly molding the resin on the sensor  \cite{POLAR-2_optical}, this thinner entrance window allowed for reducing the optical crosstalk between channels by an order of magnitude, from $\sim15\%$ to $\sim1.5\%$, thereby further improving the instrument's polarimetric sensitivity.

\begin{figure}[h]
    \centering
    \includegraphics[height=.55\textwidth]{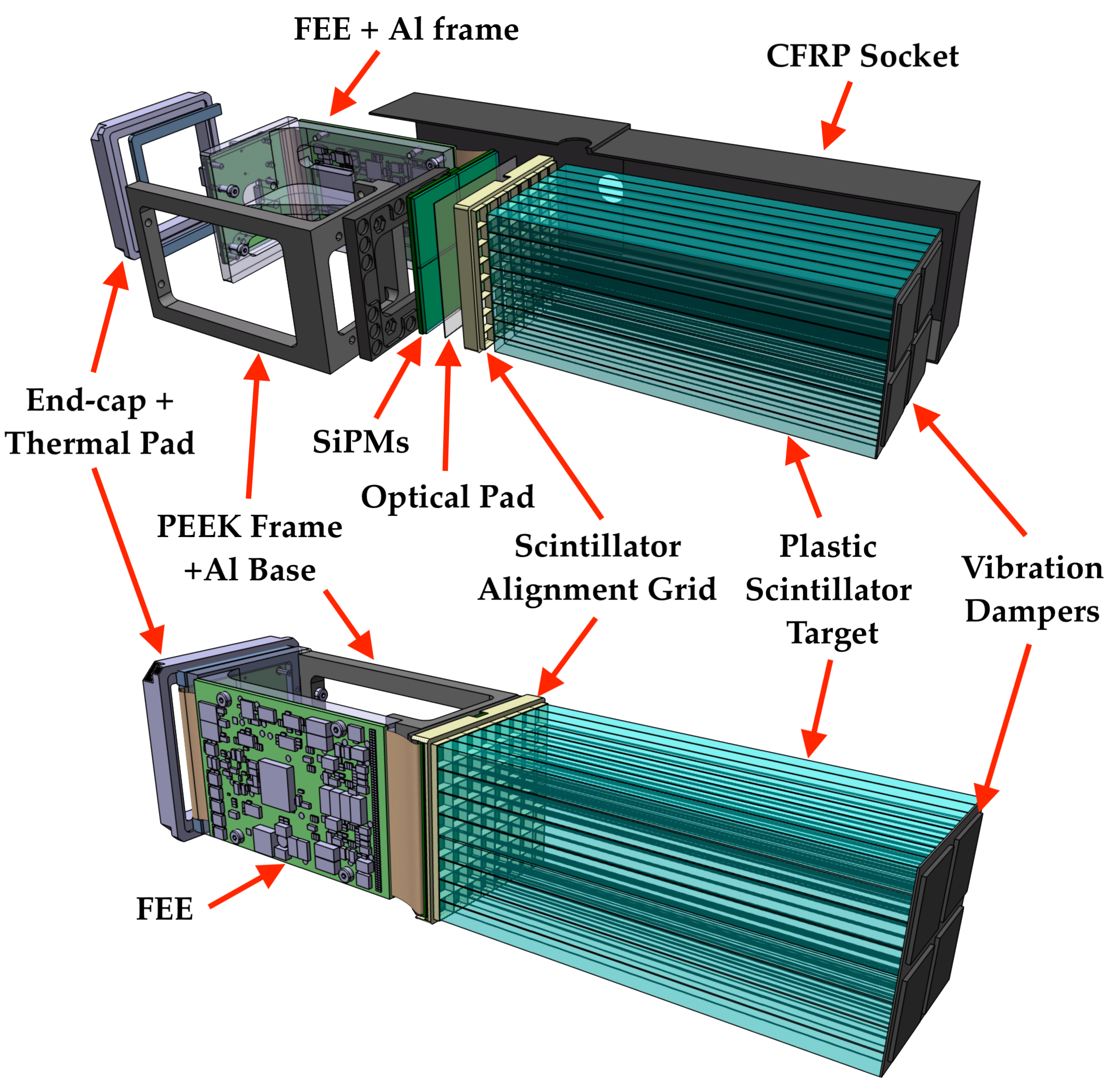}\hspace*{0.1cm}\includegraphics[height=.52\textwidth]{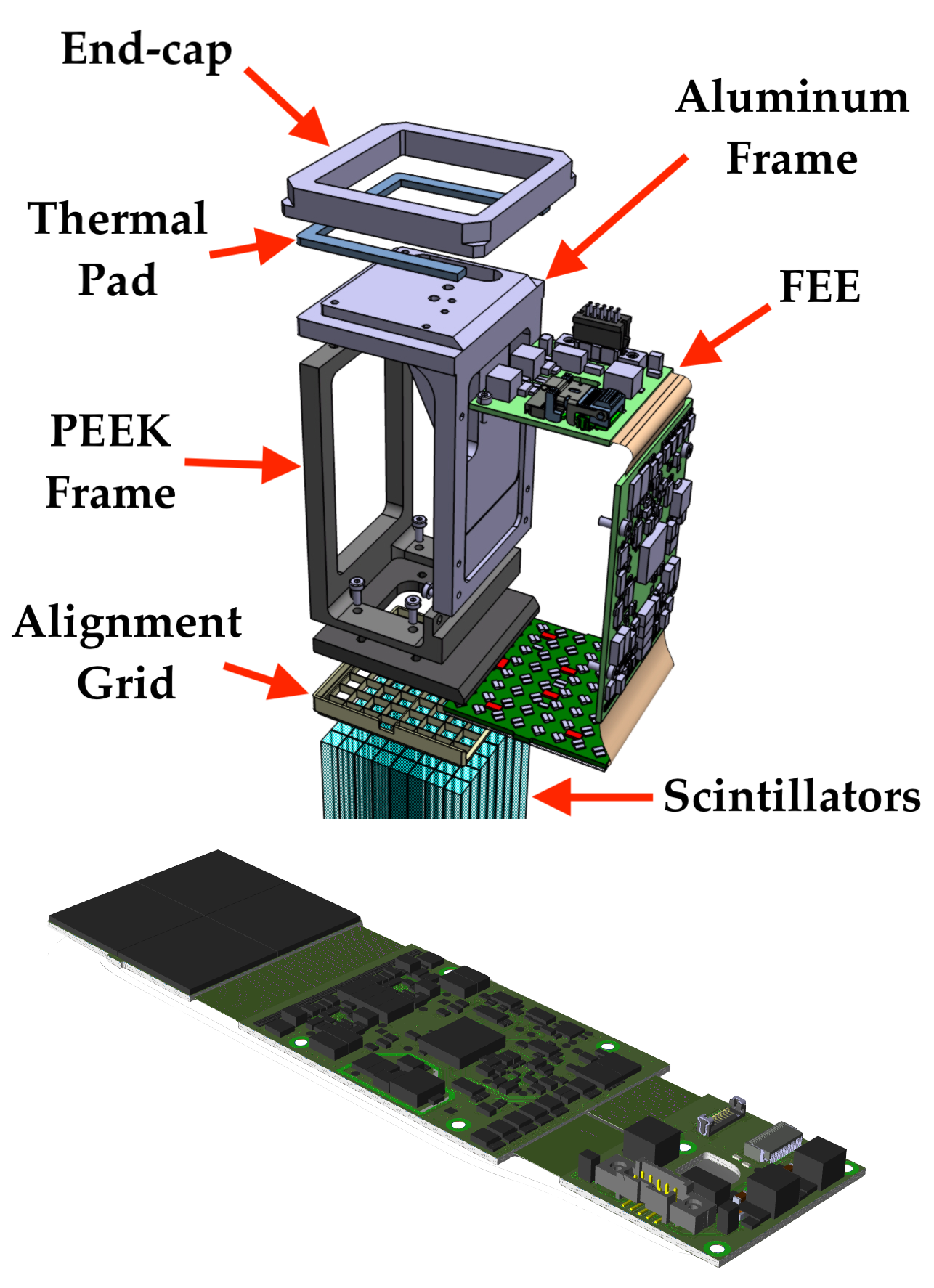}\hspace{3em}
    \caption{\textbf{Left:} Design of the HPD polarimeter module. \textbf{Top Right:} Exploded view of the module's thermo-mechanical structure optimized for heat extraction. \textbf{Bottom Right:} Design of the front-end electronics (FEE). (\textit{Module CAD images: courtesy of Franck Cadoux (UniGe/DPNC); FEE CAD image: courtesy of Yannick Favre (UniGe/DPNC)})}
    \label{fig:hpd_module_design}
\end{figure}

\subsubsection{Scintillator Material}

The reason why polyvinyltoluene (PVT) based scintillators are used despite their relatively poor scintillation efficiency and energy resolution compared to inorganic scintillators is to keep the Compton cross section dominant down to lower energies than if a high-Z scintillation material was utilized. As for POLAR, Eljen's EJ-248M plastic\footnote{\url{https://eljentechnology.com/products/plastic-scintillators/ej-244-ej-248-ej-244m-ej-248m} -- \textit{Consulted on 25\textsuperscript{th} September 2025.}} is used to convert the incoming X/$\gamma$-rays into optical light. Higher scintillation efficiency PVT-based materials, such as EJ-200 also from Eljen, were investigated to potentially increase the light yield of the system, as more scintillation light is produced in EJ-200 than in EJ-248M. A better light yield was, however, obtained with EJ-248M than with EJ-200. It was later understood through a combination of optical measurements and simulations that the surface roughness of the polished scintillator, which depends on the softness of the material, has a strong impact on the fraction of optical light getting out of the scintillator. It is indeed able to counter-balance a difference in scintillation efficiency, having a higher-scintillation efficiency material providing less light at its output due to a rougher surface. This effect is crucial for selecting the scintillation material in any high-energy experiments with a relatively low energy threshold, especially in hard X-ray polarimetry, where getting enough statistics is of great importance. A detailed discussion on this issue and implications can be found in \cite{POLAR-2_optical}. In particular, the root mean square roughness $R_q$ was measured to be $123\pm21$~nm for EJ-200 and $62.0\pm1.0$~nm for EJ-248M, the latter leading to a light yield $\sim$26\% larger despite a lower scintillation efficiency thanks to a $\sim$20\% larger light output fraction. The simulated light output fraction, that is the fraction of light that goes out of the scintillator due to interface losses, as a function of the scintillator roughness and the position of the $\gamma$-ray interaction in the scintillator vertical direction, is given in Figure~\ref{fig:roughness_annealing}. The roughness is expressed in terms of the $\sigma_\alpha$ parameter, which is the Gaussian spread in inclination of micro-facets defined in Geant4 to emulate the surface roughness \cite{POLAR-2_optical}.

\subsubsection{Radiation Tolerance for SiPMs and Damage Mitigation Strategies}

The only major drawback of SiPMs over PMTs is the presence of dark noise due to thermally induced excitations in the silicon depletion layer. The dark count rate of a SiPM is increasing with temperature, with a typical rate of the order of MHz at room temperature, implying that the sensors should be operated as cold as possible in order to reduce this noise and lower the threshold as much as possible. A custom thermo-mechanical structure was designed, as shown in the top right image of Figure~\ref{fig:hpd_module_design}, to extract the heat from the electronics towards the back of the instruments and external radiators, avoiding to inject heat towards the sensors. An active cooling system based on a Peltier element placed on the back of the SiPM arrays was originally foreseen, but thermal simulations showed that using a thermoelectric unit to cool down the sensor was actually not efficient and was injecting even more heat into the system. A fully passive thermo-mechanical structure has therefore been designed, composed of a PolyEther Ether Ketone (PEEK) frame on the SiPM side to thermally insulate the sensors from the rest of the electronics and an aluminum frame that takes the heat from the central part of the front-end electronics (FEE) out of the polarimeter module. The central part of the FEE notably contains the Field Programmable Gate Array (FPGA) and the two Application-Specific Integrated Circuits (ASICs), which respectively consume 1~W and 2$\times$0.35~W, that emit most of the heat of the electronics. The FEE is described in detail in section~\ref{subsubsec:FEE}, and has a total power consumption of about 2~W.

\begin{figure}[h]
    \centering
    \includegraphics[height=.37\textwidth]{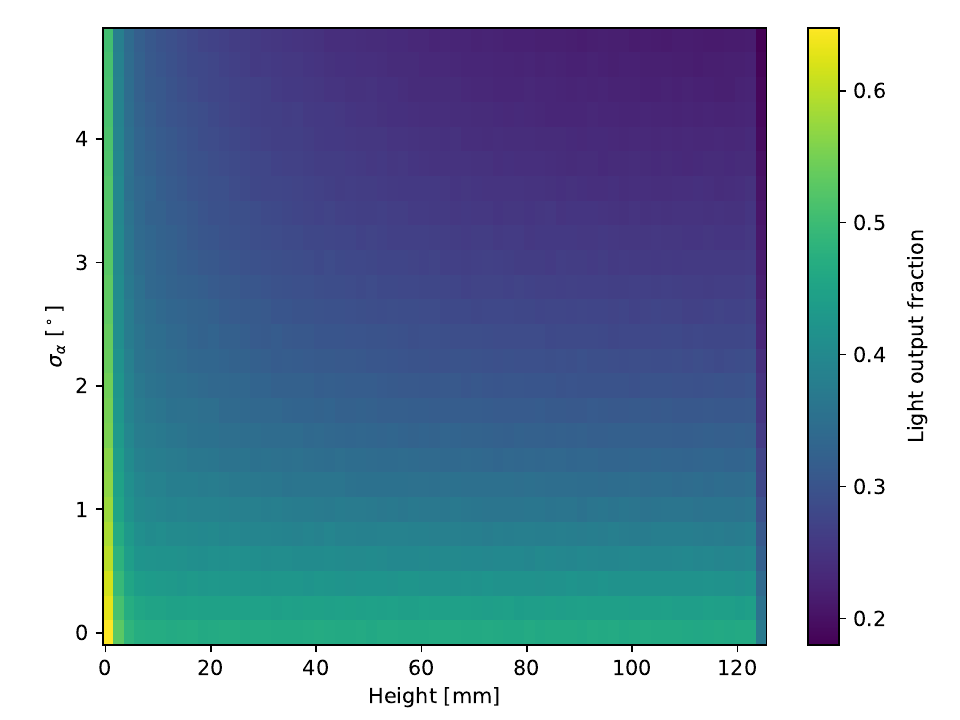}\hspace*{0.1cm}\includegraphics[height=.37\textwidth]{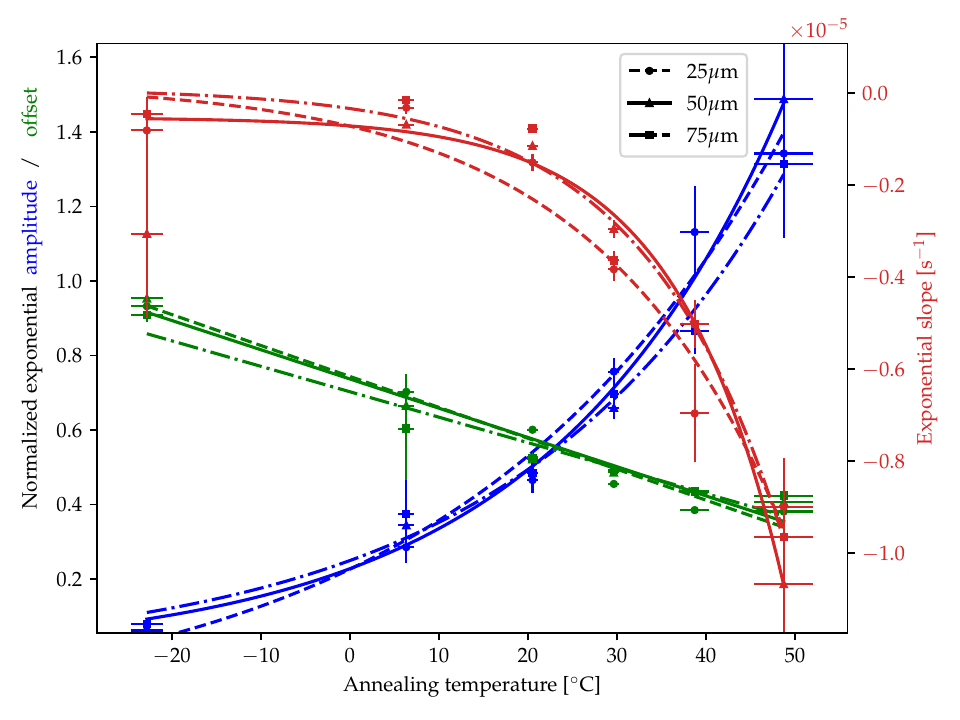}\hspace{3em}
    \caption{\textbf{Left:} Simulated light output fraction for EJ-248M as a function of the interaction point along the scintillator bar and the scintillator surface roughness (taken from \cite{POLAR-2_optical}, with permission). \textbf{Right:} Annealing-induced dark current exponential decay parameters as a function of temperature (taken from \cite{Annealing_paper}, with permission).}
    \label{fig:roughness_annealing}
\end{figure}

Another problem that arises from dark noise is the fact that it increases with radiation exposure. Indeed, radiation-induced damage in the silicon lattice causes the resistivity of the crystal to go up, thereby increasing the dark current of the sensor. This effect occurs when the sensors are exposed to a radiation environment, such as that encountered in low Earth orbit. The radiation dose rate to which the SiPMs are exposed, and therefore the amount of damage and increase in dark noise, is highly dependent on the orbit's altitude and inclination. Once in space, the instrument thus sees its dark noise increasing throughout the mission life, enforcing an increase in threshold with time. This can however be mitigated through thermal annealing of the radiation damage. Defects in the crystal can be recovered with thermal energy, implying that heating up the SiPMs results in a decrease in dark current. As this issue was a significant concern to POLAR-2, a dedicated study was performed. There the exponentially decaying dark current was characterized for the S13360 SiPM family from Hamamatsu with the three microcell pitches available on the market (namely 25, 50, and 75~$\mu$m) as a function of the sensor temperature \cite{Annealing_paper}. While the details can be found in \cite{Annealing_paper}, Figure~\ref{fig:roughness_annealing} shows the exponential decay parameters (amplitude, decay constant, and offset) as a function of temperature. A higher temperature results in both a faster recovery and a higher decrease in dark current. As a consequence, heaters have been included at the back of the SiPMs in the FEE design to perform thermal annealing of the radiation damage along the mission's operating life and partially recover the initial performances of the instrument. Several strategies have been defined as to the duration and temperature at which thermal annealing is performed \cite{Annealing_paper}, which can be applied to any space mission using similar sensors. In particular, it is foreseen during POLAR-2 operations to stop data taking for a couple of days every 6 months to heat the SiPMs at about 50$^\circ$C, which will allow to keep the energy threshold increase rate due to the increase of dark noise to a reasonable value of 0.75~keV/yr instead of about 2.2~keV/yr without performing thermal annealing.

\subsubsection{SiPM Readout Electronics System}\label{subsubsec:FEE}

A custom FEE was developed for reading out the 64 SiPM channels of each polarimeter module. In order to allow for the maximum number of readout channels in the HPD, this readout was optimized to have a low power consumption. In addition, in order to produce natural redundancy, the electronics is used for 64 channels only, and can operate largely independently. It is composed of three rigid PCB parts separated by Kapton flexes. The 3D model of the FEE can be seen in Figure~\ref{fig:hpd_module_design}. The central part contains the main parts of the electronics, while one extremity is made of four S13361-6075PE Hamamatsu SiPM arrays with RC filters for each channel, and the other contains power (3.8~V through TigerEye connector from Samtec) and signal (low-voltage differential signaling pairs through FireFly connector from Samtec) connectors as well as voltage regulators. The central part, whose dimensions are 50.8$\times$76~mm, contains over 700 electronics components. The 64 SiPMs are biased through an LT3482 DC/DC converter from Linear Technology, and their signals are digitized by two 32-channel CITIROC-1A ASICs from Weeroc. The ASIC configuration, data acquisition, and trigger logic are managed by an IGLOO FPGA from Microsemi. A local trigger logic is applied at the module level to trigger on single and double-hit events. In case of single events, the system is put on hold so that the back-end electronics can check for a coincident trigger signal from another module. High multiplicity events, which can be caused by a cosmic ray crossing the detector, are discarded. The total power consumption each readout is approximately 2 W, while its total costs are of the order of several thousand USD. This design allows for a total of 6400 channels to be operated in the HPD while keeping the power consumption below $300\,\mathrm{W}$. More details about the FEE design, performances, and potential other applications are given in \cite{Kole2025_a}.

A set of Negative Temperature Coefficient (NTC) thermistors is distributed on the back of the SiPM arrays to monitor the sensors' temperature precisely. Another NTC is placed close to the FPGA to monitor the temperature of the central FEE part. A temperature loop correction is implemented in the FPGA's firmware through lookup tables in order to update the bias voltage of each channel every second based on their breakdown voltages' evolution with temperature to keep the overvoltage constant. Non-uniformities in breakdown voltages, gain, and threshold are corrected on a channel-to-channel basis thanks to dedicated parameters in the ASIC configuration, which are characterized for each module.

\subsubsection{Overall instrument design}

The sensitive part of the polarimeter is composed of 100 modules disposed in a square array. The modules are inserted in an aluminum grid, on top of which a carbon fiber cover is placed to ensure at least 4~mm of material between the scintillators and the vacuum of space to filter the low-energy electron background. As can be seen in the exploded design of the overall HPD payload in Figure~\ref{fig:hpd_design}, the polarimeter grid is stacked on two other grids. The middle grid contains five 20-channel Low-Voltage Power Supply (LVPS) hubs that both distribute the power to every module from the LVPS and collect the signals of every module to the Back-End Electronics (BEE). The bottom grid contains the LVPS, which converts the voltages provided from the CSS on which the payload will be installed. Additionally, it hosts the BEE which controls all the system as well as the communication system responsible for transmitting data to the space station and for receiving commands from it. Finally, the instrument is foreseen to be equipped with a Beidou communication system which allows for near real-time alerts in case of a transient detection. This system is based on that employed in the GECAM instrument \cite{Beidou}.

\begin{figure}[h]
    \centering
    \includegraphics[width=.6\textwidth]{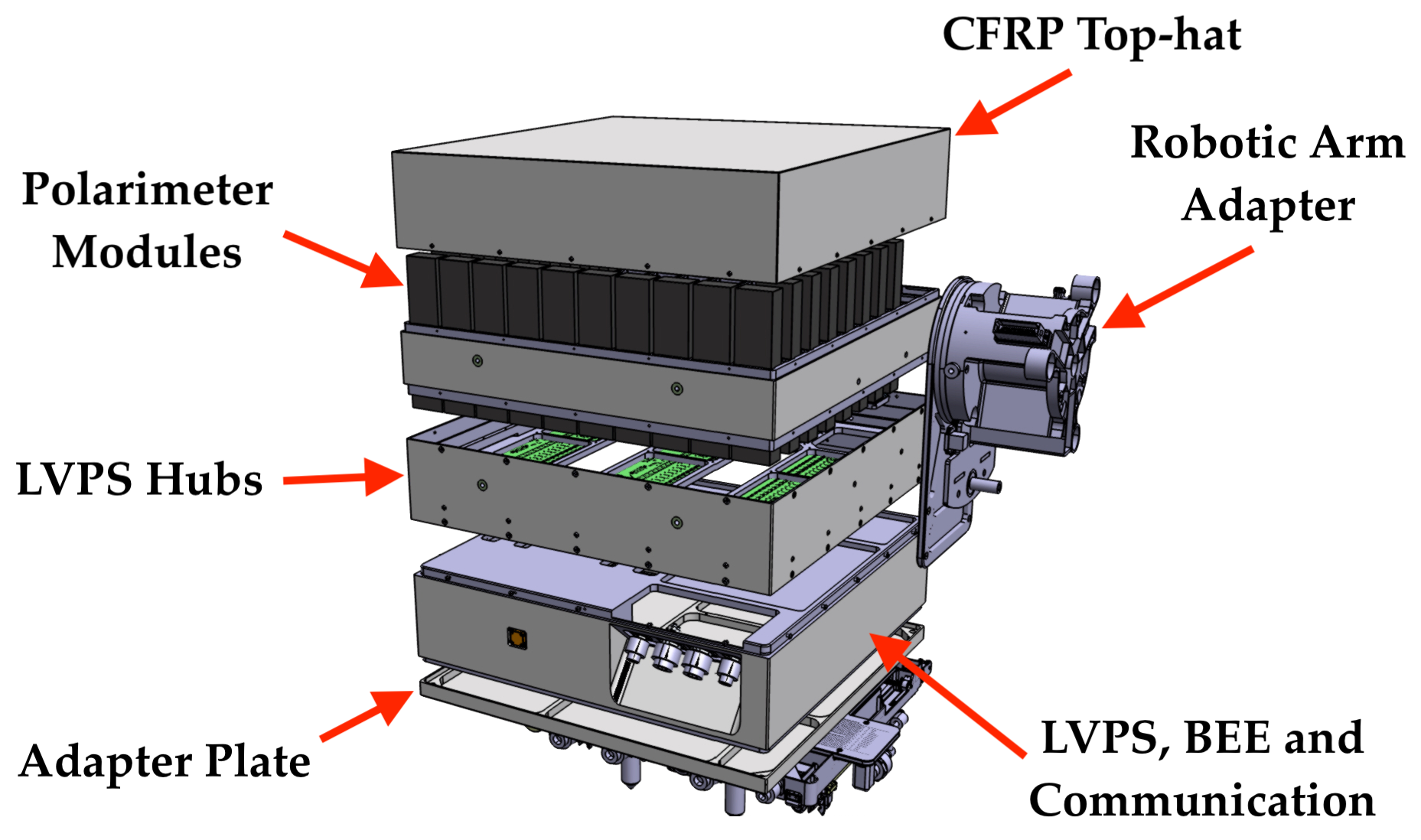}\hspace{3em}
    \caption{Exploded view of the HPD design. (\textit{Courtesy of Franck Cadoux (UniGe/DPNC)})}
    \label{fig:hpd_design}
\end{figure}

The payload, which by requirement fits in a 500$\times$600$\times$600~mm$^3$ envelope, is mounted to a platform hosting 9 scientific payloads on the CSS looking towards deep space. An aluminum robotic arm adapter is mounted on the side of the payload for installing the payload on the station once launched. This robotic arm can significantly influence the polarization signal due to signal photons scattering into the detector from this material, an effect which cannot be calibrated on ground and therefore could result in systematic errors. To mitigate this, a 10~mm-thick multi-layer shield is sandwiched between the instrument and the adapter. The exact composition of the different layers is still under study to understand the impact of fluorescence lines from the selected shielding materials on the instruments' response. The instrument will partially be covered by multi-layer insulation (MLI) while some surfaces will be exposed to space to radiate heat away from the instrument. The surfaces not covered with MLI will be coated with a multi-layer white paint from Covalba, which has been space-qualified\footnote{Tested for UV irradiation, Atomic Oxygen, and Outgassing according to ECSS standards.} for POLAR-2. The total power and mass budgets of the HPD are respectively 550~W and 160~kg (excluding the mass and power needed for the adapters).

\subsubsection{Detector Maturity and Space Readiness}

The polarimeter module design has been matured through extensive space-qualification tests. Irradiation of individual components as well as entire boards has been performed to ensure the polarimeter can survive the harsh radiation environment once in orbit, as reported in \cite{NDA_thesis, sipm_irr, scint_irr, Annealing_paper}. A polarimeter module has also been tested for sinusoidal and random vibrations, as well as shock, to validate its hardness to extreme launch conditions. A load spectrum measured in between the vibration and shock tests performed along the direction orthogonal to the scintillators is plotted in Figure~\ref{fig:hpd_spacequalif}, while more details can be found in \cite{NDA_thesis, SPIE25_POLAR-2}. A mini-polarimeter made of 3$\times$3 modules, from which 5 were thermal dummies emulating the same power distribution, was thermal cycled for about a week to cross-check the behavior of the instruments' thermal simulations crucial to understand the SiPM operating temperature once in space, which has a strong impact on the scientific performances of the instrument. This campaign also allowed to show the correct behavior of the polarimeter modules over a wide range of temperatures. The temperature monitored near the SiPMs and FPGA on three FEEs during this test is plotted in Figure~\ref{fig:hpd_spacequalif}.

\begin{figure}[h]
    \centering
    \includegraphics[height=.37\textwidth]{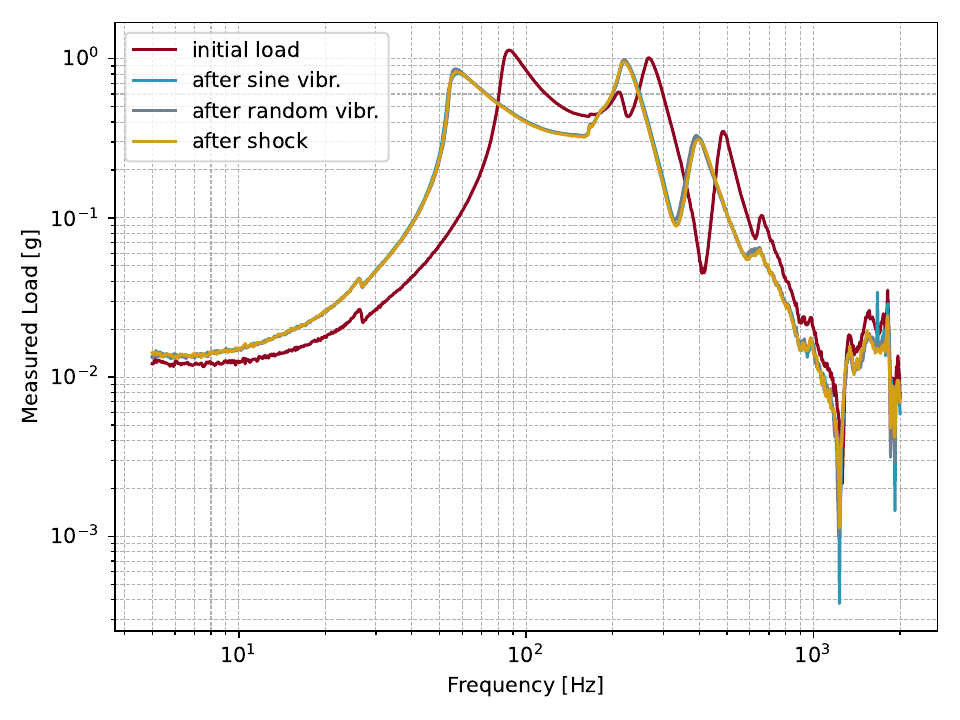}\hspace*{0.1cm}\includegraphics[height=.37\textwidth]{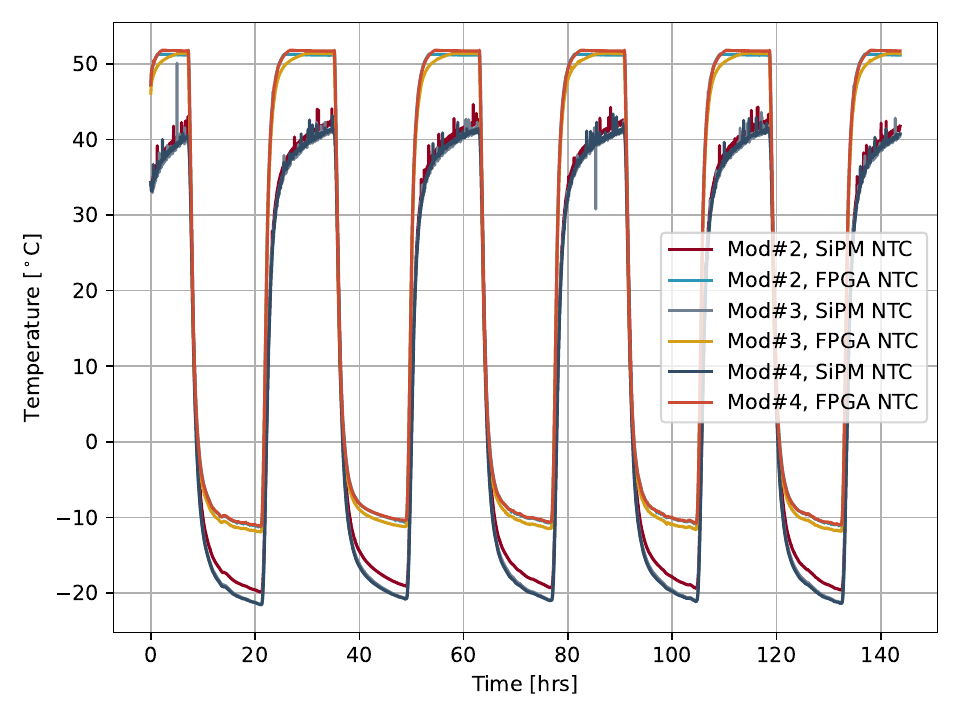}\hspace{3em}
    \caption{\textbf{Left:} Load measured, through an accelerometer placed at the base of the module next to the dampers, between each vibration and shock run along the x-axis -- in the sensor plane (taken from \cite{SPIE25_POLAR-2}, with permission). \textbf{Right:} Temperature monitored next to the FPGA and on the back of the SiPM arrays during thermal vacuum cycling of three polarimeter modules (taken from \cite{SPIE25_POLAR-2}, with permission).}
    \label{fig:hpd_spacequalif}
\end{figure}

The instrumental performances based on the first polarimeter modules built in the laboratory were assessed through a combination of simulations and calibration measurements using polarized hard X-rays, such as the beam of the European Synchrotron Research Facility (ESRF) in France where several modules were calibrated \cite{ESRF2023}, as detailed in section~\ref{sec:HPD_main}.

\subsection{SPD Design}\label{sec:BSD}

In addition to the HPD, the POLAR-2 mission also contains the SPD payload. The SPD comprises two types of detectors: a Broad-band Spectrometer Detector (BSD) and the LPD. The BSD is primarily used for providing precise GRB spectral and location information, while also possessing some polarization capabilities for bright and spectrally hard GRBs. The LPD is primarily used for measuring the polarization of GRBs in the soft X-ray band. The combination of the SPD and HPD will enable the simultaneous observation of GRB polarization, spectrum, and light curve in a wide energy range. The BSD will also be able to provide a sub-degree level self-localization result for GRBs, which helps to improve the GRB polarimetry samples for POLAR-2. 

\subsubsection{Broad-band Spectrometer Detector}

Analysis of the data from POLAR-2's predecessor, the POLAR experiment onboard of ``Tiangong-2'', indicated the importance of having information on the GRB incoming direction and spectrum as input conditions for the polarization analysis. This is a result of the polarization response not only being strongly dependent on the polarization, but also on the energy spectrum and incoming angle. For example, the $M_{100}$ value depends on both these parameters. As a result, an uncertainty on these, results in an inaccuracy in the calculation of the PD and therefore induces a systematic error in the polarization measurements \cite{Kole2020}. Since POLAR's self-localization accuracy was of the order of several degrees, and the plastic scintillator material used in the detector was not able to provide high-precision spectral measurement results in the concerned energy detection range, these pieces of information often needed to be provided by dedicated GRB spectrometers. Although polarization measurements were performed for several GRBs where such information was missing, the higher precision analysis results were all thanks to simultaneous observations of POLAR by either \textit{Fermi}-GBM or Swift-BAT \cite{Kole2020}. Thus, the number of POLAR GRB polarization observation samples depends to some extent on whether these GRB instruments can provide accurate localization results and spectral measurement parameters at the same time.

For the polarization measurements with the HPD and LPD, the design of the BSD requires that for those GRBs detected by these instruments a localization can be measured. For example, to maintain the systematic uncertainty from localization below $\sim$1\%, a GRB with a fluence of $10^{-5}$ erg/cm\textsuperscript{2} in the 10--1000~keV range must be localized to within $\sim1^\circ$ of its true location. Based on this requirement, the BSD's detector localization capability design considers using the tungsten coded-aperture mask imaging method. While meeting the core performance requirements of the BSD, to enhance engineering feasibility and reduce costs, the cerium-doped $Gd_{3}Al_{2}Ga_{3}O_{12}$ (GAGG:Ce) scintillation crystal was selected. This was chosen based on its high light yield of $\sim45$ optical photons per keV, non-hygroscopic nature and mechanical stability \cite{Kole2025_a}. SiPM devices are employed to collect and process the optical signal from GAGG:Ce crystals, allowing for a BSD development with a shorter development cycle, lower technical risk, and more controllable overall budget.

The BSD instrument mainly consists of a tungsten coded-aperture mask plate and an imaging detector array. The imaging detector comprises 36 Detector Modular Units (DMU), forming a 6x6 array. Each DMU contains 64 GAGG crystal bars, forming an 8x8 array, and the same FEE as used on the HPD \cite{Kole2025_a}. The mask is composed of $\sim$3500 tungsten alloy elements (each measuring 6.25 mm $\times$ 6.25 mm $\times$ 1.00 mm), arranged in a random pattern with a 50\% open fraction (Fig.~\ref{fig:mask_design}). To minimize weight, the mask plate is reinforced with low-density carbon fiber composite layers, each $\sim$1.50 mm thick. The overall dimensions of the mask are 599 mm $\times$ 499 mm, with an active imaging area of 575 mm $\times$ 475 mm. To obtain more GRBs jointly observed by HPD and LPD, the field of view (FOV) of BSD is designed to be as large as possible. The fully-coded field of view (FCFOV) is approximately $80^\circ \times 54^\circ$, and the half-coded field of view (HCFOV) is about $132^\circ \times 125^\circ$. There are 2304 GAGG crystal bars in total composing the BSD detector array, and the dimensions of each GAGG crystal bar are 5.75 mm $\times$ 5.75 mm $\times$ 20 mm, making the total detection area about 760 cm\textsuperscript{2}. The front-end electronics of BSD will adopt the same design as HPD, with each piece of SiPM independently corresponding to the signal readout of GAGG crystal bar. Through tests of a prototype it was found that the current design can perform spectral measurements in the $\sim10-1000\,\mathrm{keV}$ energy range. On board of the FEE, the FPGA is used to control the ASIC to read out the SiPM signals. The FEE's high integration and low power consumption facilitate the use of large-area detector arrays. Furthermore, the BSD is designed with an in-orbit trigger and localization function intended to quickly transmit GRB trigger and localization information in orbit down to the ground using the CSS's platform, guiding ground-based and other satellite observation instruments to conduct follow-up observations, achieving multi-mission and multi-wavelength collaborative observations, and increasing the scientific output of the entire project. A comprehensive description of the BSD instrument's design is provided in a companion paper currently in preparation \cite{BSD_paper}.

\begin{figure}[h]
    \centering
    \includegraphics[width=0.6\textwidth]{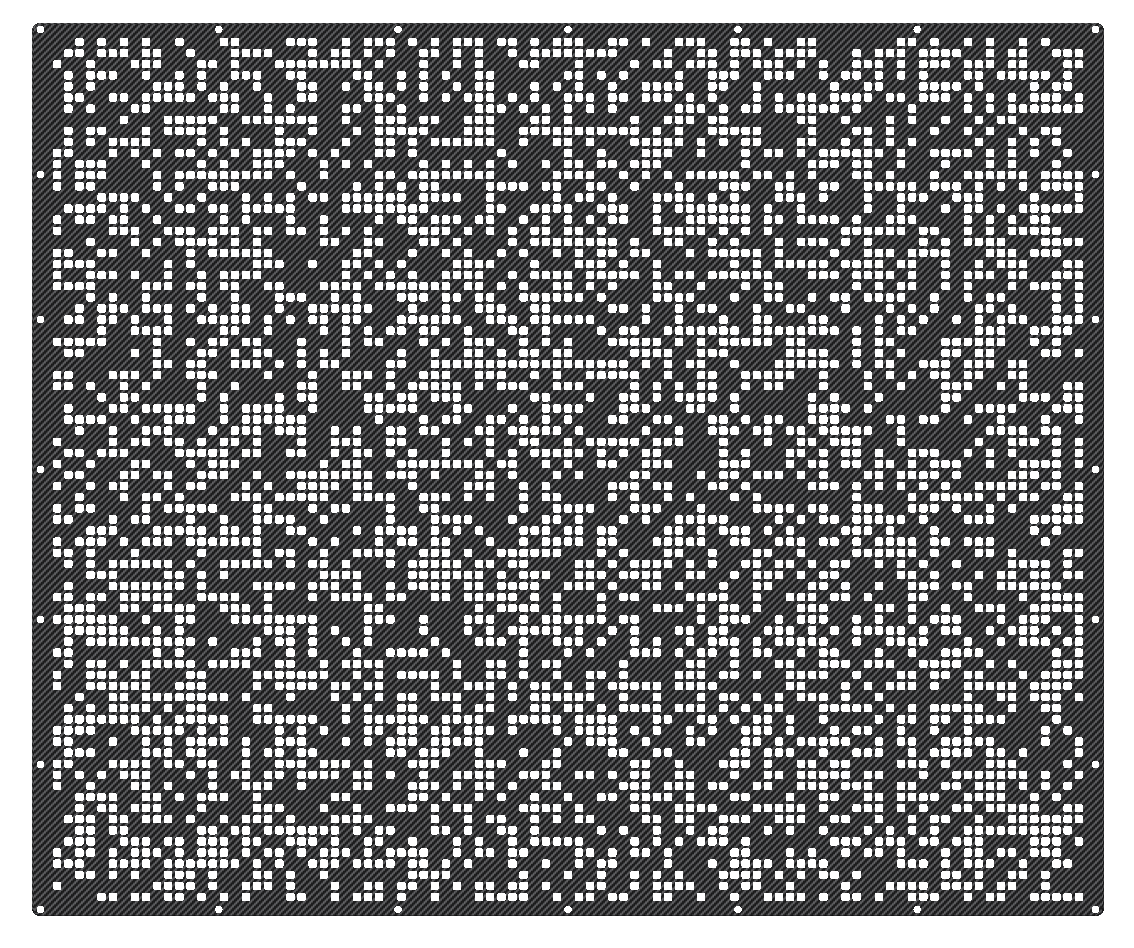}\hspace{3em}
    \caption{Design of the coded-aperture mask plate.}
    \label{fig:mask_design}
\end{figure}

\subsubsection{Low-energy Polarization Detector}

The primary goal of the LPD is to measure the soft X-ray polarization of GRBs in the 2–10 keV energy range. The main detector material is a gas mixture of helium (He) and dimethyl ether (DME). The measurement relies on tracking photoelectrons generated when incident X-rays interact with the gas through the photoelectric effect, as their emission directions carry information about the polarization state of the incoming photons. Each LPD detector modular unit consists of a gas chamber, a Gas-based Micro-Channel Panel (GMCP), and a multi-pixel ASIC chip, Topmetal-L \cite{2025JInst..20P7009A}. The LPD detector, shown in Figure \ref{fig:lpd}, comprises three LPD detector modules, and each module contains three Gas Micro-Channel Panel Detector (GMPD) units \cite{Feng,Feng2} along with a dedicated readout electronics system, resulting in a total of nine gas detector units. The total polarization-sensitive detection area of the LPD is approximately $30\,\mathrm{cm^2}$, with a field of view of about $90^\circ\times90^\circ$ \cite{Yi:2024keb}. The implementation of LPD will enable the first wide-field survey of soft X-ray polarization in the 2–10 keV range.

\begin{figure}[h]
    \centering
    \includegraphics[width=0.7\textwidth]{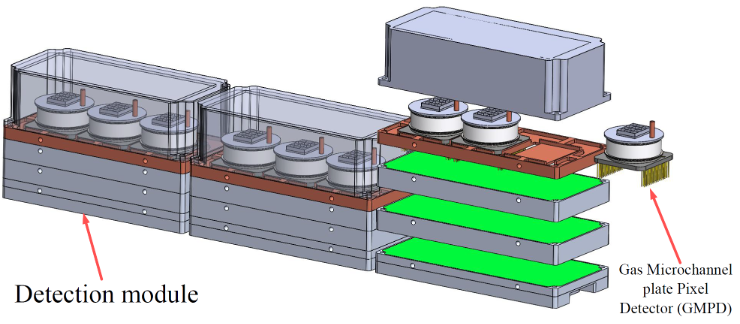}\hspace{3em}
    \caption{Design of the LPD payload.}
    \label{fig:lpd}
\end{figure}

\begin{figure}[h]
    \centering
    \includegraphics[width=0.7\textwidth]{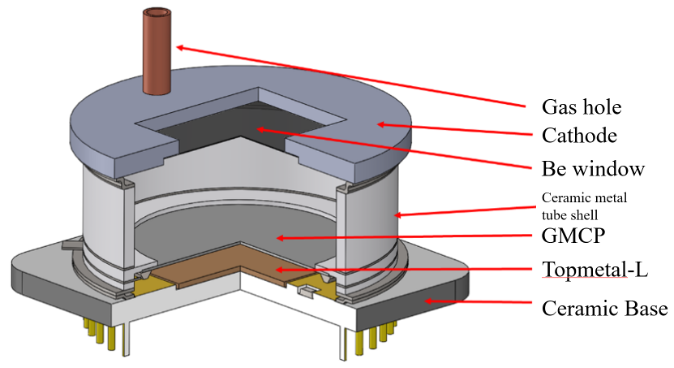}\hspace{3em}
    \caption{Prototype design of the GMPD.}
    \label{fig:GMPD}
\end{figure}

The prototype of the GMPD detector consists of a cathode, a GMCP, a Topmetal-L pixel readout chip, a ceramic metal tube shell, and a ceramic base, as shown in Figure \ref{fig:GMPD}. Among these components, the cathode, metal–ceramic tube shell, and ceramic base form the sealed chamber of the detector. The cathode section is composed of a Kovar alloy frame, a beryllium (Be) window, and copper tubes. To ensure that X-rays can effectively pass through the window while maintaining airtightness and mechanical reliability of the detector structure, the beryllium window, with a thickness of 100~$\mathrm{\mu m}$, is positioned at the center of the cathode. The Be window is tightly bonded to the Kovar frame using a mature brazing technique, which ensures vacuum sealing as well as the required structural strength. Copper tubes are arranged along the edge of the cathode and serve critical functions for vacuum pumping and gas filling. These tubes are also brazed to the Kovar frame for a firm connection. After the working gas is filled into the chamber, the openings of the copper tubes are sealed using ultrasonic welding technology. The ultrasonic welding machine provides a highly reliable seal for the copper tubes, ensuring the integrity of the detector chamber.

The ceramic metal tube shell adopts a sophisticated structural design, in which three ceramic rings and four Kovar alloy rings are hermetically joined through brazing technology. Each pair of adjacent Kovar rings is separated by a precisely positioned ceramic ring. These ceramic layers not only provide electrical insulation but also ensure accurate spacing between the Kovar layers. This structural configuration guarantees both the overall mechanical strength of the shell and the required electrical insulation performance.

At both ends of the tube shell, the Kovar rings are hermetically sealed to the cathode and the ceramic base using laser welding, forming a robust and airtight chamber. This connection method ensures the structural integrity of the shell while providing a solid foundation for the installation of the GMCP. The two Kovar rings located in the middle of the shell are specially designed to hold the GMCP and serve as its output electrodes, providing the necessary electrical paths for electron multiplication and signal transmission.

The ceramic base is also composed of ceramic and Kovar alloy. A Kovar ring is mounted on the top of the base and is laser-welded to the corresponding Kovar ring on the metal–ceramic shell to achieve a hermetic seal. To ensure uniformity of the electric field in the induction region, a rectangular slot is machined at the center of the ceramic base to accommodate the Topmetal-L chip, ensuring that the chip surface is level with the base surface. Additionally, the inner surface of the ceramic base is copper-coated and grounded to promote uniform electric field distribution. The signal pads of the chip are wire-bonded to the corresponding pads on the ceramic base.

To monitor the working gas conditions inside the GMPD, a BME680 pressure and temperature sensor is mounted on the ceramic base. The sensor module adopts a compact metal-lid LGA package and is a digital sensor based on a mature sensing principle, featuring small size, low power consumption, and the capability to measure both gas pressure and temperature. Both data transmission and power supply for the Topmetal-L chip and the gas sensor are realized through pins integrated into the ceramic base.
Based on the above design, a single GMPD detector unit has dimensions of $50\times50\times40\mathrm{mm^3}$. The gap between the cathode and GMCP forms the drift region with a height of 14~mm, while the gap between the GMCP and the anode forms the induction region with a height of 1~mm. The detection area of one detector is $3.691\mathrm{cm^2}$. All components requiring hermetic sealing are joined using welding techniques—mainly brazing and laser welding—to ensure high gas tightness and excellent mechanical performance of the gas detector.

The Gas-based Micro-Channel Plate (GMCP) is made of an insulating substrate coated with conductive metal layers on both sides, serving as electrodes. It consists of an array of microscopic channels—thousands of fine pores—through which electron multiplication occurs. Within the GMCP, electrons are accelerated by the applied electric field and gain sufficient kinetic energy to ionize gas molecules or atoms, generating secondary electrons. These newly produced electrons are further accelerated and collide with additional gas molecules, initiating a cascade of secondary ionizations that result in an electron avalanche amplification process.

In terms of fabrication, the GMCP shares similarities with traditional Micro-Channel Plates (MCPs). It uses lead–bismuth silicate glass as the base material. The manufacturing process involves drawing glass fibers, stacking fiber bundles, grinding and polishing, chemically etching to form microchannels, and using a hydrogen reduction process to tune the bulk resistance. Nickel–chromium alloy electrodes are deposited on both ends of the GMCP through evaporation, ensuring strong adhesion and good electrical conductivity. The GMCP surface is exceptionally smooth and nearly free of burrs, so no additional polishing is required before use.

Furthermore, through hydrogen reduction treatment, a conductive layer is formed inside each channel, imparting a controlled bulk resistivity to the GMCP which is shown in Figure \ref{fig:GMCP}. This property helps dissipate accumulated charges and effectively suppresses charge buildup effects, thereby significantly enhancing the stability of gas detectors \cite{Feng3}. Therefore, the detector would be ideal for observing transient sources where the flux of particles changes rapidly. Additionally, the GMCP material exhibits a very low outgassing rate, which minimizes contamination of the working gas by released materials and consequently extends the detector’s operational lifetime.

\begin{figure}[h]
    \centering
    \includegraphics[width=0.7\textwidth]{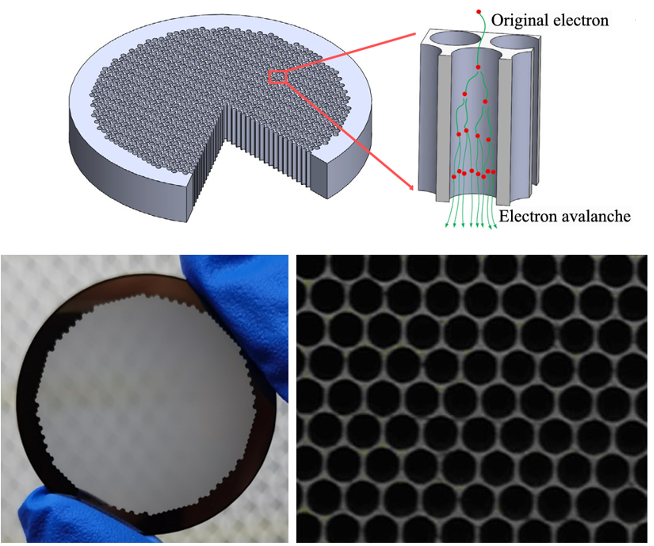}\hspace{3em}
    \caption{\textbf{Top:} A schemtic overview of the full GMCP as well as the working principle of a single channel. \textbf{Bottom:} Images of the GMCP to be used in the LPD.}
    \label{fig:GMCP}
\end{figure}

The readout is performed using the Topmetal-L which is a CMOS pixel sensor chip specifically designed for the LPD. Its design is based on the previously developed Topmetal-II \cite{TM1, TM2} and Topmetal-M \cite{TMM} chips from Central China Normal University. Fabricated using a 130~nm CMOS process, the chip adopts an innovative pixel scanning architecture—known as the Sentinel Readout Scheme which differs from the traditional rolling shutter readout method. Compared with the latest Topmetal-M2 chip, Topmetal-L significantly reduces the number of output channels from 16 to 1 and the number of readout buffers (Abuffers) from 32 to 1, thereby greatly lowering power consumption. As a result, the chip achieves a total power consumption of less than 1 W. A schematic diagram of the Topmetal-L structure is shown in Figure \ref{fig:Topmetal}.
The Topmetal-L chip measures $17\times24\mathrm{mm^2}$ and contains a 356$\times$512 pixel array with individual pixel dimensions of $45\mathrm{\mu m}\times45\mathrm{\mu m}$. Peripheral circuitry is located along the left and bottom sides of the pixel matrix, while the chip’s I/O pads are evenly distributed along the left, right, and bottom edges. Internally, the chip also integrates L-shaped level shifters, BIAS voltage circuits, and driver buffers, with the L-shaped digital scan control readout module positioned at the outermost lower-left corner.

\begin{figure}[h]
    \centering
    \includegraphics[width=0.5\textwidth]{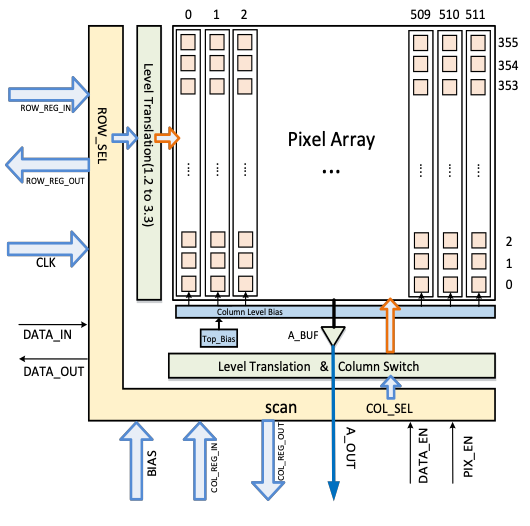}\hspace{3em}
    \caption{\textbf{Top:} Schematic Diagram of the Topmetal-L Chip Structure}
    \label{fig:Topmetal}
\end{figure}

\subsubsection{Overall Payload}

In addition to BSD and LPD, the SPD payload also includes a Back-End Electronics (BEE) system used for the data transmission, power supply, and configuration of BSD's GAGG detectors and LPD detectors, and it is also responsible for interfacing with the CSS. The four sidewall panels of the SPD payload provide shielding and mechanical support, reducing the main in-orbit background influence on GRB localization. The shielding panels consist mainly of 0.4 mm thick tantalum (Ta) and a combination of aluminum (Al) alloy ranging from 1 to 10 mm thick.

The overall design of the SPD payload is shown in Figure~\ref{fig:spd_design}.

\begin{figure}[h]
    \centering
    \includegraphics[width=1.0\textwidth]{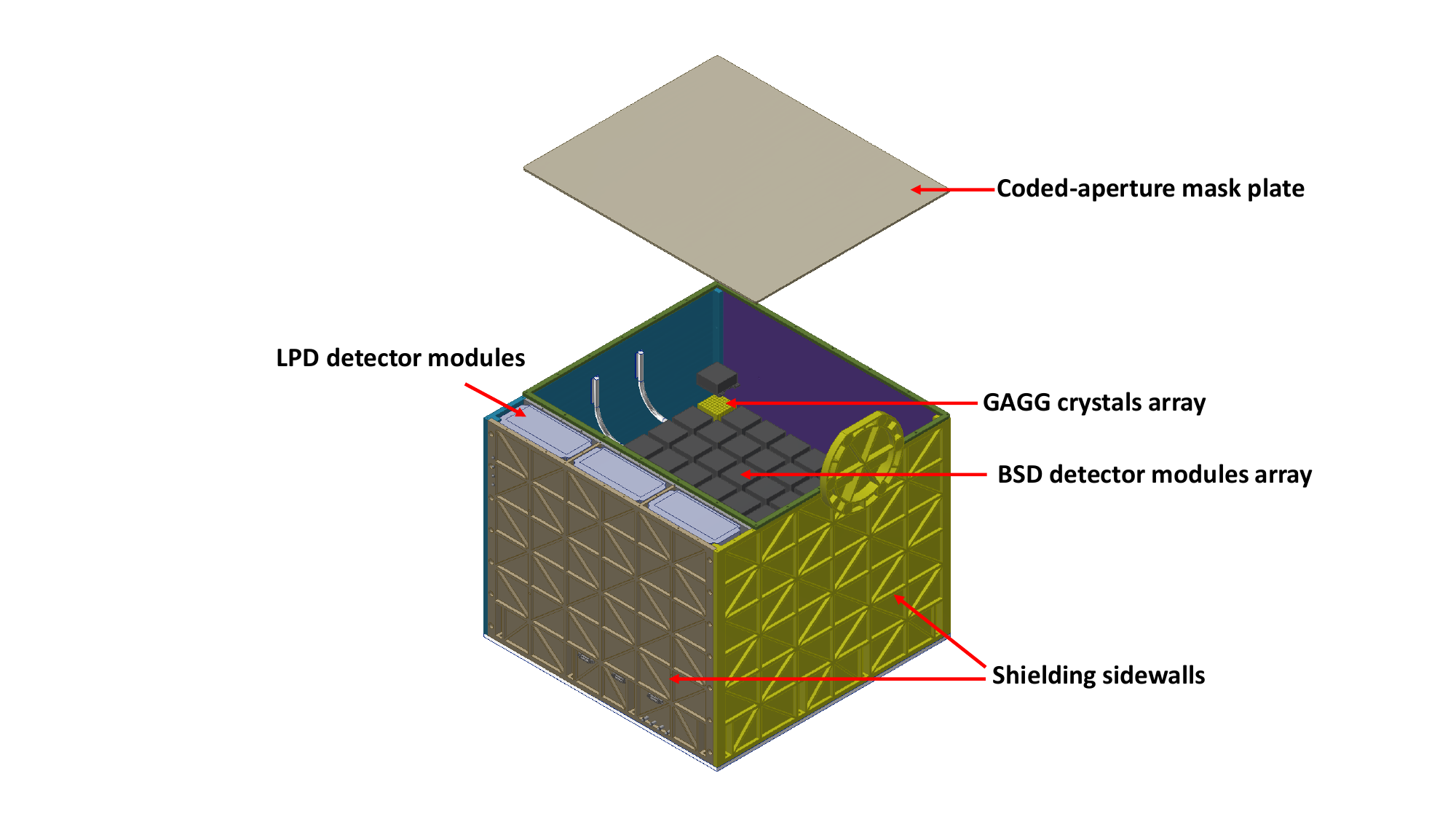}\hspace{3em}
    \caption{Design of the SPD payload.}
    \label{fig:spd_design}
\end{figure}

\section{Measurement Prospects}\label{sec:Prospects}

\subsection{HPD Performance}\label{sec:HPD_main}

The first prototypes of detector modules of the HPD have undergone extensive testing using both radioactive sources and polarized beam facilities. Such tests mainly serve to verify the simulated instrument performances produced using the Geant4 framework \cite{Agostinelli2003GEANT4}. The most recent measurements of this type for the HPD were presented in \cite{ESRF2023}, where a single detector module was subjected to polarized beams in the $40-120\,\mathrm{keV}$ energy range. As presented in \cite{ESRF2023} the simulations match the measurements well and therefore provide confidence when simulating the response of the full HPD instrument. For this purpose, a detailed mass model of the instrument was simulated using a range of both polarized and unpolarized photons of various energies. Similar to how this was successfully done for POLAR, as discussed in \cite{Kole2020}, Geant4 was only used to simulate the physical interactions and to output the energy depositions and the locations of the photon interactions. The produced output file was subsequently processed using a secondary simulation software which handles the conversion from energy, to optical photons to electrical signals and handles the processing of these signals in the electronics. Using these two steps, the total effective area of POLAR-2, which includes also triggers which cannot be used for polarization analysis, was produced and is shown in Figure~\ref{fig:ARF}. The effective area shown here is for an on-axis source. In addition, the effective area simulated for events which contribute to the polarization measurements was produced, as shown on the right of Figure~\ref{fig:ARF}. The total effective area can be seen to peak close to $2500\,\mathrm{cm^2}$ while the polarization effective area exceeds $1000\,\mathrm{cm^2}$ above 100 keV.

By simulating both an unpolarized and a polarized flux for this GRB the $M_{100}$ can also be calculated as a function of energy. The result of this is also shown in Figure~\ref{fig:M100} for both an on-axis and a $30^\circ$ off-axis source. It should be noted here that these results were produced using basic assumptions on the threshold to be set to a minimum achievable ($\sim5\,\mathrm{keV}$) and without any cuts on the distance between the interaction positions. Studies where these values are further optimized are still to be performed. For example, an increase in the detection threshold will of course reduce the effective area of the instrument but will increase the $M_{100}$ and in addition, reduce the background rate. For the POLAR mission optimization studies allowed the $M_{100}$ to peak at $\sim45\%$ for an on-axis source at around 200~keV as can be seen from the public POLAR instrument responses \footnote{\url{https://www.astro.unige.ch/polar/grb-light-curves}}.

While optimization studies are still required, comparing the response of POLAR-2 to its predecessor, it is clear that a significant gain in the effective area is achieved. This improvement is most prominent at energies below 100~keV. Due to the lower detection threshold both the $M_{100}$ and the effective area are significant starting from $\sim40\,\mathrm{keV}$ compared to $\sim80\,\mathrm{keV}$ for POLAR. This can be understood as the threshold allows to detect scattering interactions of photons of these energies as illustrated on the right side of Figure~\ref{fig:M100} which shows the energy deposited in a $180^\circ$ Compton scattering interaction as a function of the incoming photon energy. The mean energy threshold of POLAR corresponded to the back-scattering of an $\sim80\,\mathrm{keV}$ photon, implying that it is not possible to detect any Compton scattering interactions, and therefore to perform any polarization, below these energies.  

\begin{figure}
    \centering
    \includegraphics[width=0.44\textwidth]{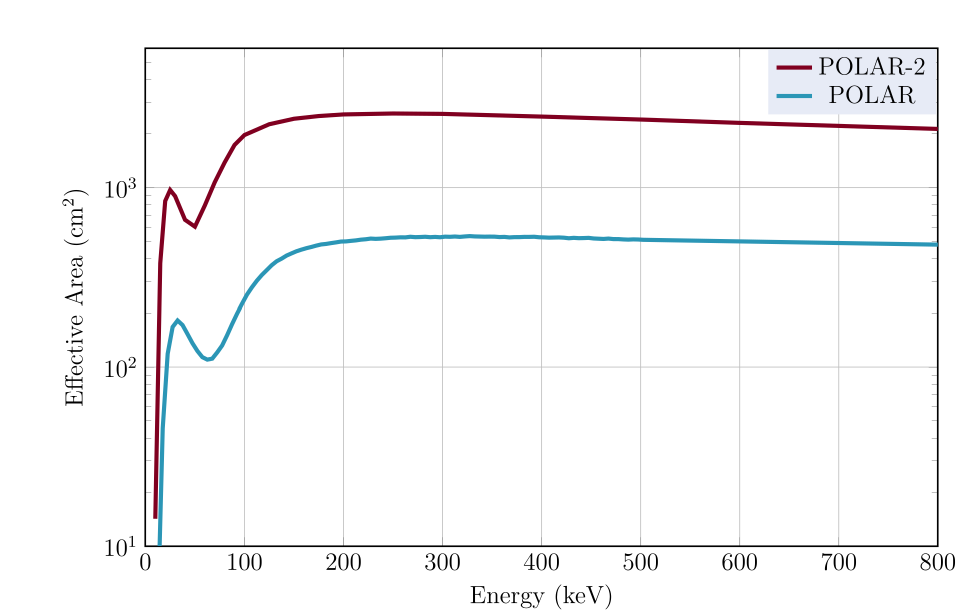}    \includegraphics[width=0.45\textwidth]{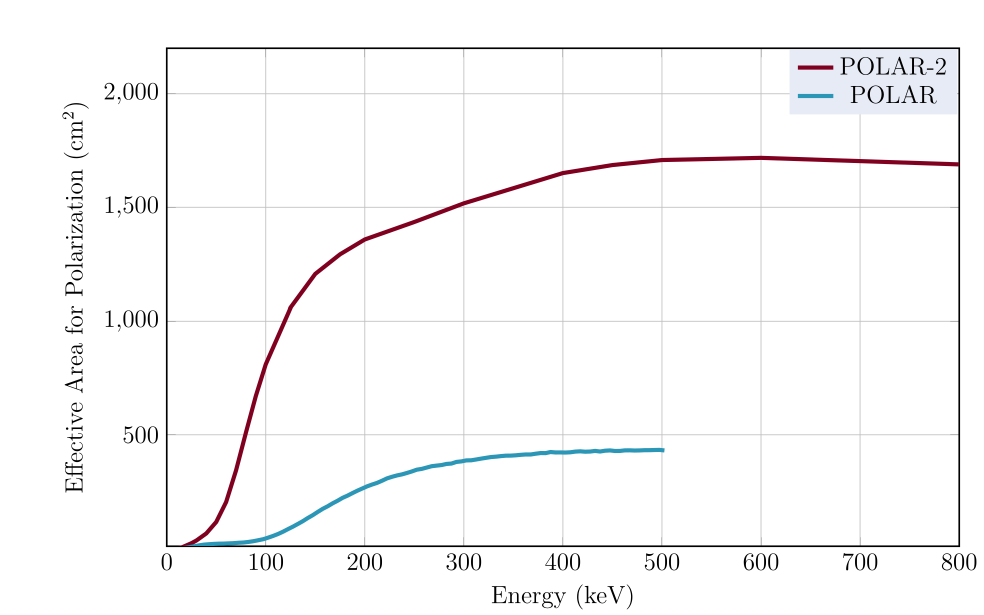}
    \caption{\textbf{Left:} The total effective area of POLAR-2 compared to that of POLAR. The effective area shown here includes all valid triggers. \textbf{Right:} The effective area for polarization events only of POLAR-2 compared to that of POLAR. The dynamic range for polarization events for POLAR typically ended around $500\,\mathrm{keV}$ }
    \label{fig:ARF}
\end{figure}

\begin{figure}
    \centering
    \includegraphics[width=0.45\textwidth]{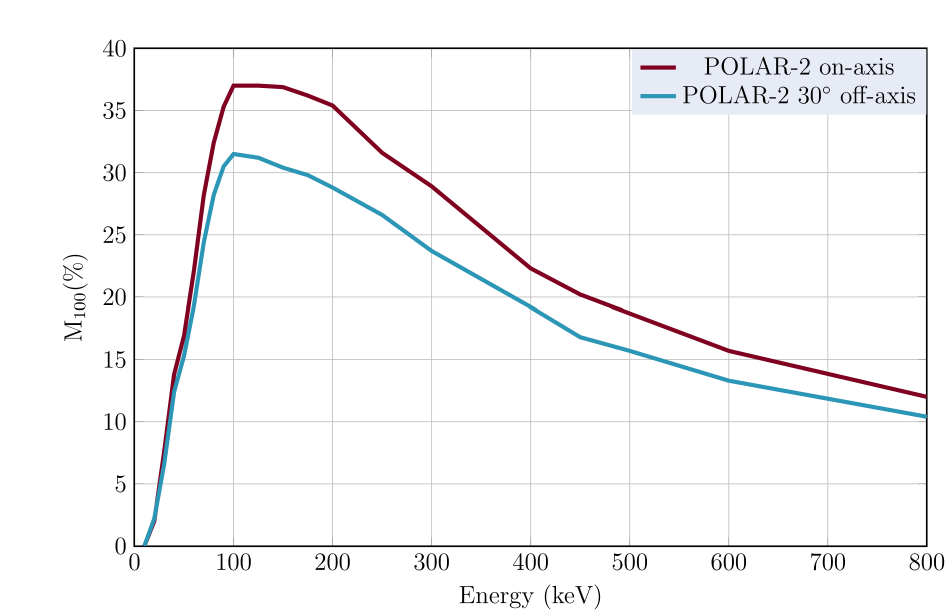} 
    \includegraphics[width=0.45\textwidth]{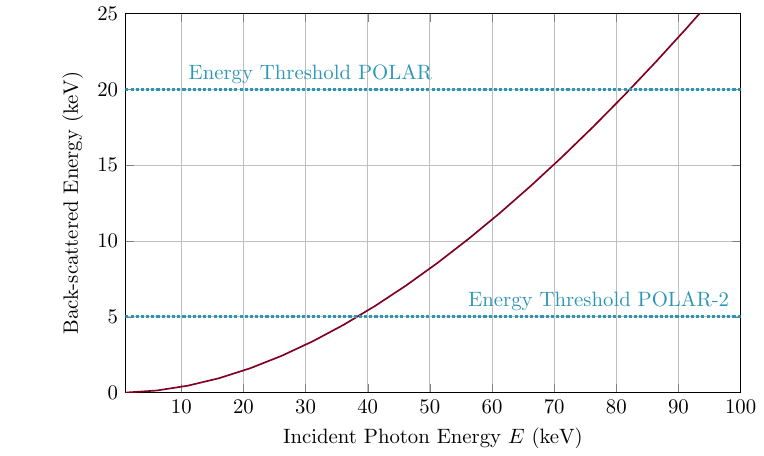} 
    \caption{\textbf{Left:} The modulation factor as a function of energy of POLAR-2 for both and on-axis and a $30^\circ$ off-axis source. \textbf{Right:} The maximum energy from Compton scattering (scattering at $180^\circ$) as a function of the photon energy. The energy mean threshold of POLAR and that achievable with POLAR-2 are indicated with the horizontal lines to indicate at which energy polarization measurements become possible.}
    \label{fig:M100}
\end{figure}

The effective area and the $M_{100}$ of POLAR-2 as a function of energy were simulated for various incoming $\theta$ angles. While the effective increases with $\theta$, before reaching a maximum around $30^\circ$, the $M_{100}$ peaks at $\theta=0$. An example of the $M_{100}$ as a function of energy for $\theta=30^\circ$ is shown on the left of Figure \ref{fig:M100}. Although the field of view of POLAR-2, using its definition from spectrometry, is half the sky, efficient polarimetry measurements become difficult for objects which exceed $70^\circ$ as the $M_{100}$ starts to peak below $15\%$. This, in combination with a decreasing effective area starts to affect the MDP significantly. However, for particularly bright GRBs, constraining polarization measurements are still possible at such angles. Rather than stating a polarimetry equivalent of the field of view, we here study in detail the number of GRBs for which POLAR-2 can perform accurate polarization measurements.

In order to make a realistic prediction on the MDP values for GRBs observed by POLAR-2 during its operation time, we took the public HEASARC GRB detection catalog from \textit{Fermi}-GBM\footnote{\url{https://heasarc.gsfc.nasa.gov/w3browse/fermi/fermigbrst.html}} \cite{vonKienlin2020} and used their spectral parameters, energy integrated photon flux and $T_{90}$ time, along with a randomized position on the sky, to simulate a GRB catalog observed by the POLAR-2 mission. For this, the GRB observed by \textit{Fermi}-GBM between the start of 2015 and the end of 2019 were used and their number was reduced by a factor of $30\%$ to account for the difference in the amount of sky observed by \textit{Fermi}-GBM and POLAR-2. The spectra from these GRBs were combined with a random $\theta$ angle, which allowed to select the effective area and $M_{100}$ for that specific $\theta$. The spectra were then folded through the response to produce the number of counts as well as the spectral averaged $M_{100}$ (where the energy-dependent $M_{100}$ is weighted for the number of detected photons in each energy bin). This, in combination with an estimated background rate of $2000\,\mathrm{Hz}$ and the $T_{90}$ of the GRB allows to calculate the MDP. It should be noted that this estimated background rate is a value extrapolated from the measurements of POLAR which had a background rate of $\sim500\,\mathrm{Hz}$. 

The background of POLAR was dominated by the cosmic X-ray background with a small additional contribution from Earth albedo photons and neutrons \cite{PRODUIT2018259}. Despite not having significant shielding, the background induced by charged particles was negligible. For low-energy charged particles this was due to passive shielding by at least 4 mm of carbon fiber for each detector element which stops the majority of charged particles with energies below 2 MeV. The effectiveness of this was demonstrated by the relatively uniform distribution of the background among the detector elements in POLAR. A significant contribution from charged particles would have resulted in a significantly higher count rate in the scintillator bars located closest to outer space, as these provide additional shielding to the scintillators closer to the center. However, no such increase was observed \cite{PRODUIT2018259}. High-energy charged particles were instead vetoed online using the deposited energy in a single scintillator or through the multiplicity of the number of triggered scintillators. Whereas gamma-rays will typically only trigger 2 or 3 detector channels, charged particles will produce ionizing tracks while traversing a large number of scintillators. In addition, a Minimum Ionizing Particle (MIP) will also deposit at least 1.2~MeV in a single scintillator. In POLAR the front-end electronics therefore vetoed any events where the total deposited energy in a single detector module was above a set level as well as events where more than 7 detector channels produced a trigger. Based on this success, for POLAR-2 a similar strategy will be employed while the exact settings will be optimized prior to launch. Given the lower optical crosstalk, and the extended dynamic range, such settings need to be optimized compared to those used on POLAR.

Despite such possible improvements, here we use an increase of a factor of 4 of the background for POLAR-2 compared to POLAR. This is based simply on the increase in number of readout channels and should therefore be considered an approximation. The rate is likely to increase due to the lower energy threshold, while we expect it to decrease due to the shortened scintillator bars. While a more detailed set of simulations is required to produce detailed estimates, we here assume that these two effects will cancel out and therefore keep an increase of a factor 4.

\begin{figure}
    \centering
    \includegraphics[width=0.45\textwidth]{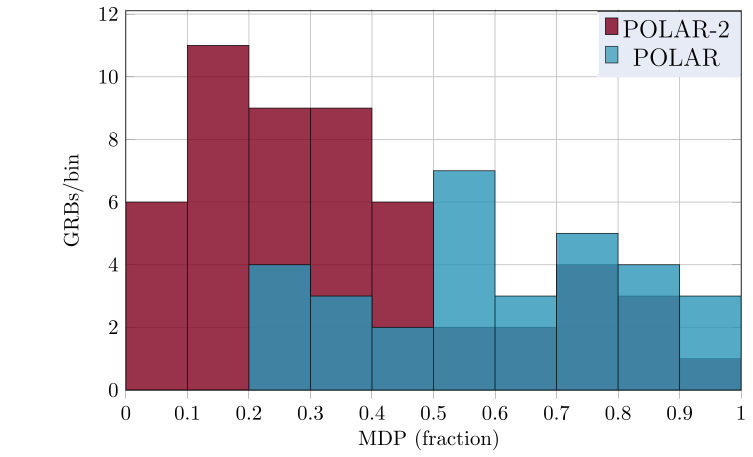} 
    \includegraphics[width=0.45\textwidth]{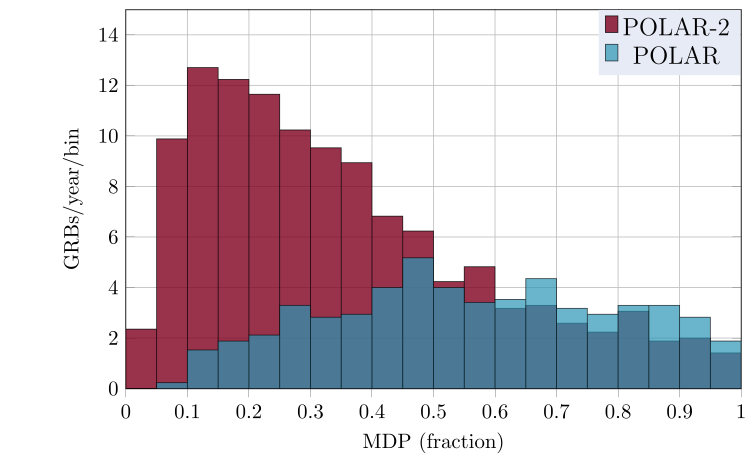} 
    \caption{\textbf{Left:} The distribution of the MDP value for the GRBs observed by POLAR (green) and POLAR-2 (red) for a mission duration equal to POLAR which is approximately half a year. \textbf{Right:} The same as the left but now using 9 years of \textit{Fermi}-GBM data to produce the yearly distribution.}
    \label{fig:M100_comp}
\end{figure}

In order to check whether the analysis procedure is accurate, it was first applied to the POLAR response using the GRB data from the period during which POLAR was active. The simulated distribution of the MDP for all the GRBs observed by POLAR during the half year is shown in blue on the left of Figure~\ref{fig:M100_comp}. Since the incoming angles are randomized here, the distribution is not equal to the real MDP distribution observed by POLAR, however, they are consistent. We subsequently applied it to the HPD of POLAR-2 to produce the red histogram on the left in Figure~\ref{fig:M100_comp} which indicates that the number of GRBs observed with an MDP below $10\%$ is significant even during half a year of observation. Then, using 9 years of \textit{Fermi}-GBM data to calculate the yearly expected rate instead for both instruments, the results presented on the right side of Figure~\ref{fig:M100_comp} were produced.

Although Figure~\ref{fig:M100_comp}  indicates a significant improvement in the number of GRBs detected with a low MDP, it does not directly tell us how well POLAR-2 can achieve its science goals. We therefore used the same simulation framework to simulate the PD values measured by POLAR-2 given either a photospheric emission scenario or a $B_{tor}$ emission model. For both the distributions of the intrinsic PD were taken from their prediction distributions as shown in Figure~14 from \cite{2020MNRAS.491.3343G}. This implies a PD centered around $\sim50\%$ for the $B_{tor}$ model while the photospheric model typically predicts a PD close to $0\%$. The true PD was provided, along with the spectrum and the $T_{90}$ and a random $\theta$ angle to calculate a probability distribution for measuring a certain PD. For each GRB, a PD was then picked randomly from this distribution. For this study we first assumed the observation period equal to that of POLAR and show the distribution for both models as expected by this instrument. This is shown in the top left of Figure~\ref{fig:model_comp}. It can be seen that although, different distributions are observed, they still largely overlap. It should be noted that the real measured distribution by POLAR looks similar to that of the photospheric case with most measured PDs being close to 0. However, several outliers with a best fitting PD exceeding $50\%$ exist. Although the POLAR results, like these simulations, therefore allow to give a strong hint that $B_{tor}$ models can be excluded a strong claim cannot yet be made. 

If we replace POLAR by the HPD, but keep the observation time to be about half a year, we get the two distributions as shown in the top right of Figure~\ref{fig:model_comp}. Although we now see two clearly different distributions, a significant number of outliers still causes overlap between the two. However, as the number of GRBs in the distribution is significant we can apply a cut an only keep bright, or those with a small $\theta$ in the distribution. This is achieved by only selecting GRBs from this distribution with an MDP below $30\%$. The results is seen in the bottom left of Figure~\ref{fig:model_comp}. Here we start to see a clear difference between the distributions which would certainly allow to rule out one of these two emission models. We can therefore conclude that the sensitivity of POLAR-2 is sufficient to differentiate between these two models within half a year of launch. To indicate what the distributions would look like after 2 years of observations we produced the bottom right of Figure~\ref{fig:model_comp} where also the cut on the MDP was increased to only include those with an MDP below $15\%$. We there see that after 2 years a very clear distinction between the models is evident.

\begin{figure}
    \centering
    \includegraphics[width=0.45\textwidth]{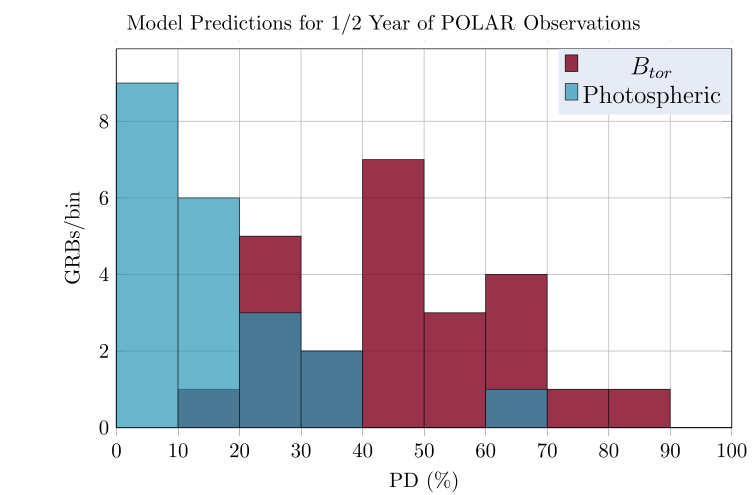} 
    \includegraphics[width=0.45\textwidth]{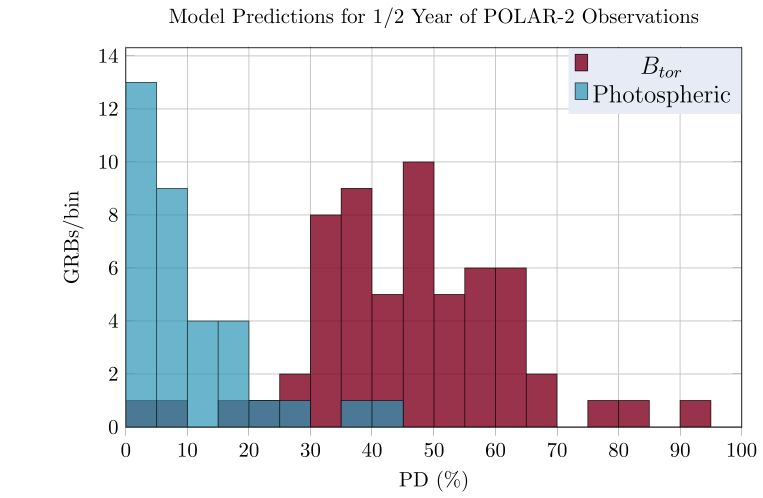} 
    \includegraphics[width=0.45\textwidth]{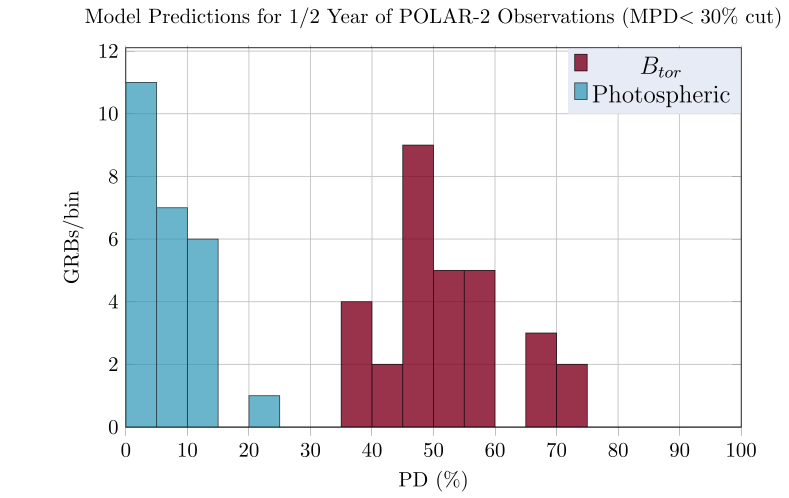}
    \includegraphics[width=0.45\textwidth]{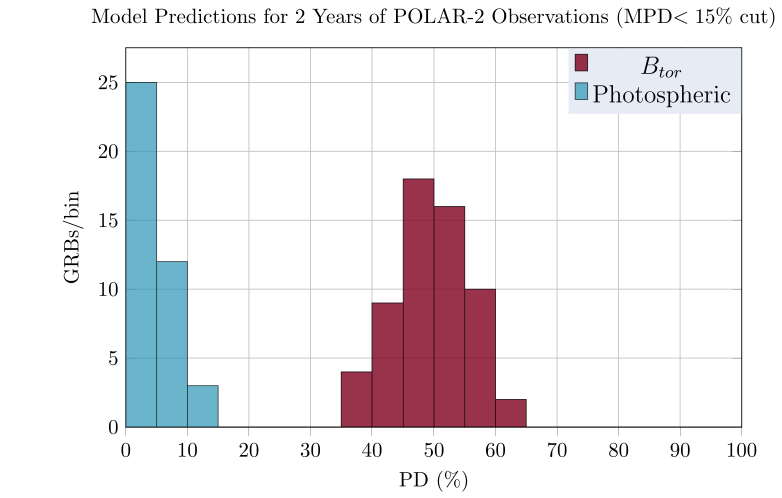} 
    
    \caption{\textbf{Top Left:} The predicted distributions of the measured PD for 2 different GRB emission models as predicted for the POLAR mission for its observational time of half a year. \textbf{Top Right:} The same as the top left but now for the POLAR-2 mission. \textbf{Bottom Left:} The same as the top right, but when only selecting GRBs with an MDP below $30\%$. \textbf{Bottom Right:} The predicted distributions as observed for POLAR-2 but now for 2 years of observation time with a cut on the MDP of $15\%$.}
    \label{fig:model_comp}
\end{figure}

From the above discussion it is evident that POLAR-2 will within several months of operation, be capable of either confirming or disproving $B_{tor}$ synchrotron models. If however, the PD distribution is more consistent with that of the photospheric model it will be more complicated to further distinguish between the remaining emission models such as Compton Drag or synchrotron models with random magnetic field structures. Although such models do predict different levels of PD, their predicted distributions are all close to $0\%$ \cite{2020MNRAS.491.3343G}. In addition, the exact shape of these distributions depends on a variety of other unknown parameters such as the jet opening angles, their Lorentz factors and the jet structure. Further distinguishing between the remaining models, if $B_{tor}$ models can be ruled out, therefore requires more complex studies than time and energy integrated polarization measurements.

\subsection{Time Resolved Polarimetry with the HPD}

The simulations used to produce the predictions of section~\ref{sec:HPD_main} consisted of unpolarized simulations in the 0 to 750~keV energy range as well as the same simulations but for a $100\%$ polarized beam with a PA of $0^\circ$. Studying the performance of POLAR-2 for time-resolved studies, specifically when studying a varying polarization angle, requires an instrument response for all possible PAs. Such a response consists of 3D grid of modulation curves for different incoming photon energies, polarization degrees and polarization angles and can be used in forward folding analyses such as those available in the \textit{ThreeML} framework \cite{threeml}. A similar set of polarization responses was used for the POLAR analysis to produce the results such as those presented in \cite{Burgess2019} for time-resolved studies of 170114A and in \cite{Kole2020} to study the full GRB catalog as measured by POLAR. Here, we produce one such response for POLAR-2 for an incoming off-axis angle of $\theta=26.4^\circ$ and $\phi=4.9^\circ$. However, rather than producing the response using only the simulation framework, which typically requires several weeks of simulations on a cluster, the procedure as discussed in \cite{Zuke, Zaid} was used to produce the modulation curves for different PAs based only on the simulated unpolarized response.

The response produced in this manner was used in combination with the \texttt{threeML} framework to study the capability of POLAR-2 to perform time-resolved analysis of GRB 170114A. This GRB, with a fluence of $1.9\times10^{-5}\,\mathrm{erg/cm^2}$ was observed by POLAR and Fermi-GBM \cite{GCN_170114A}, due to its brightness and relatively small off-axis angle of $\theta=26.4^\circ$, resulted in the highest sensitivity measurement from the POLAR catalog. As a result, as well as because of its Fast Rising Exponential Decay (FRED) like nature, the GRB was used to perform a detailed time-resolved spectral polarization analysis in \cite{Burgess2019}. This study found that, when dividing the GRB into 8 periods based on a Bayesian blocks method, the PD was most consistent with $30\%$ while the PA appeared to vary significantly during the pulse. Although the PD in each bin was still consistent with a PD of $0\%$, the results sparked significant interest and several papers discussing a theoretical interpretation of the measurements were published \cite{170114A_int1, 170114A_int2}. In case the PA evolution consists of sharp changes of $90^\circ$ the results can be explained by several theoretical models. In case the PA does, instead vary more gradually, the results are difficult to interpret \cite{Gill2021}. As pointed out in \cite{Gill2021}, a proper interpretation of GRB 170114A would require a more precise measurement of the PD as well as the PA. 

\begin{figure}
    \centering
    \includegraphics[width=0.6\textwidth]{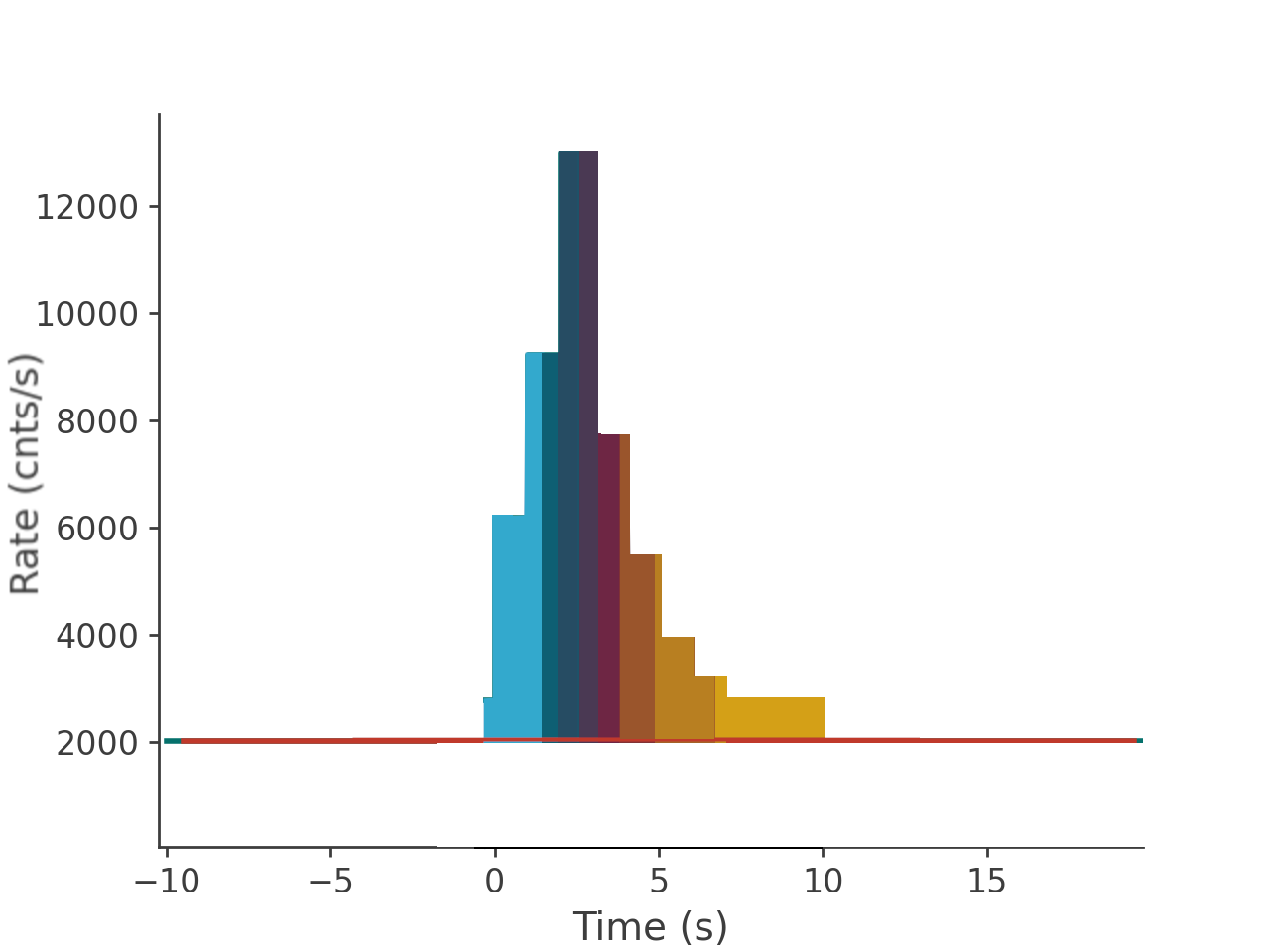}\hspace{3em}
    \caption{The light curve of 170114A based on the brightness and spectra measured by POLAR and Fermi-GBM folded through the POLAR-2 response. The background rate in POLAR-2 here was scaled by a factor of 4 compared to that of POLAR.}
    \label{fig:170114A_lc}
\end{figure}

Here, we folded the measured brightness and spectrum of GRB 170114A through the POLAR-2 response while the PD was set to $30\%$ with initially a smoothly changing PA during the first 3 time bins, followed by 2 flips of $90^\circ$. The resulting light curve is presented in Figure~\ref{fig:170114A_lc}. The background here was set to $2000\,\mathrm{Hz}$ which is approximately 4 times that of POLAR. The time bins as used in the original study with POLAR data are indicated with the different colors in Figure~\ref{fig:170114A_lc}. Using the same time bins and analyzing the simulated data with \texttt{threeML} we get the polarization measurements for the 8 time bins as presented in Figure~\ref{fig:time_res_p2}.

\begin{figure}
    \centering
    \includegraphics[width=0.8\textwidth]{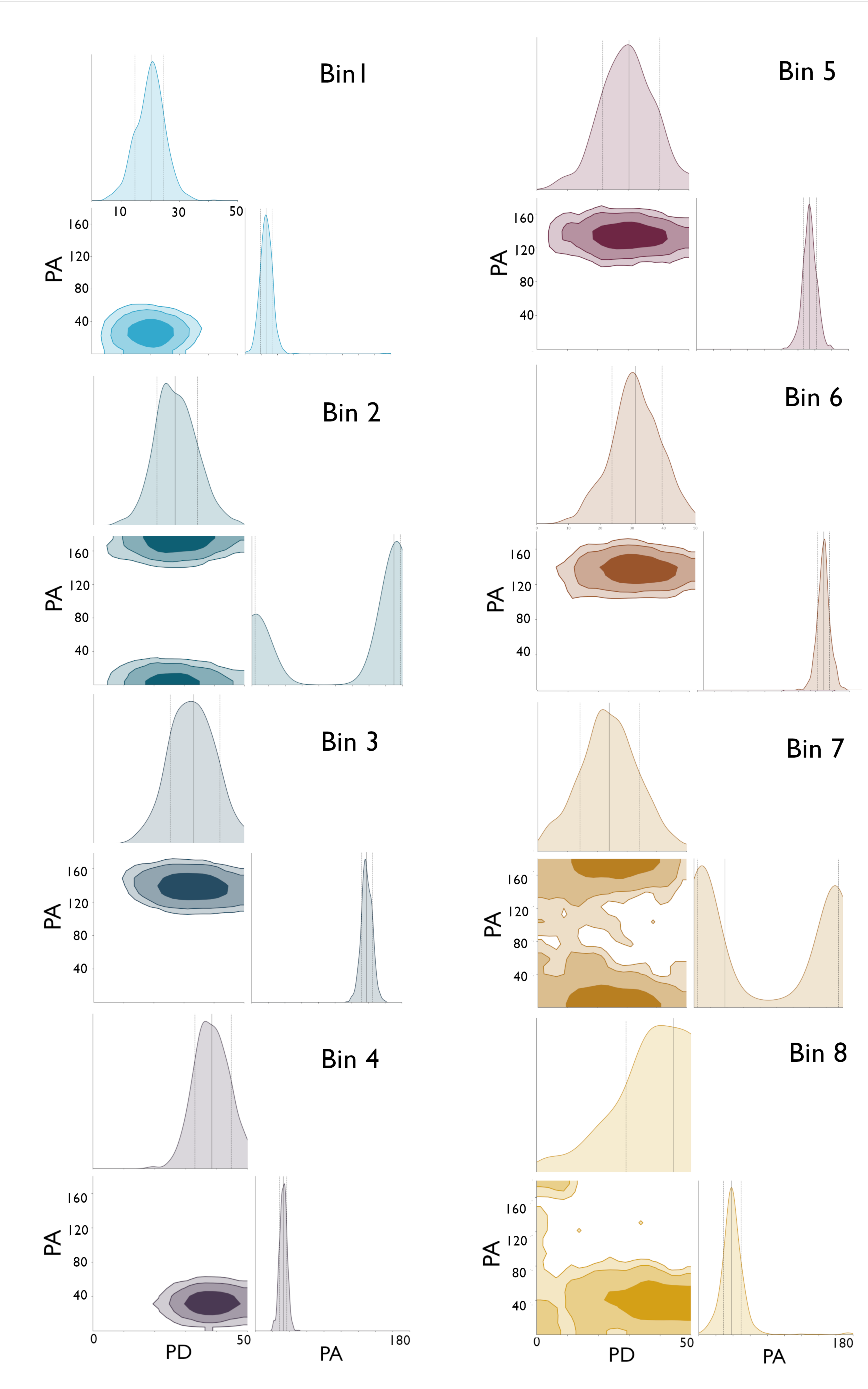}\hspace{3em}
    \caption{The posterior distributions of PD and PA for the 8 time bins resulting from the time-resolved analysis of POLAR-2 for a 170114A-like GRB with a constant PD of $30\%$ and an evolving PA. The PD is shown in the $0-50\%$ range, while the angle is shown over the full range. The middle gray line indicates the best fit position, while the lines on either side of it indicate the area which contains $68\%$ of the posterior.}
    \label{fig:time_res_p2}
\end{figure}

We see firstly that for the first 6 time bins, the PD measurement is constrained and an unpolarized emission can be excluded with over $99\%$ confidence. In addition, both the $90^\circ$ flips can be observed well (from bin 3 to 4 and from bin 4 to 5). We can therefore conclude that if POLAR-2 were to observe a GRB such as 170114A, of which based on Figure~\ref{fig:M100_comp} it will see several per year, it will be able to perform time-resolved analysis which will allow to confirm the hints of an intrapulse evolution of the PA by POLAR. In addition, the observed variations in the PA can, in combination with the PD distributions allow to further distinguish between emission models.

\subsection{Energy Resolved Polarization Measurements}

Model discrimination can also be performed through energy-resolved studies. While few models exist on the energy dependence of the polarization of the prompt emission, one can expect that such dependencies should exist. A prime example of this is when the emission consists of dissipated photospheric emission at high energies, and synchrotron emission at lower energies such as in \cite{Lundman_2018}. There the dependence of the PD for such an emission model was calculated specifically for GRB 990123. Based on the results presented there, we plot the predicted PD for such a GRB on the left side of figure \ref{fig:Lundman_plot} where the energy ranges in which the HPD and the LPD are sensitive are highlighted as well. We can see that while the PD is significant in the energy range of the LPD whereas it is negligible in the majority of the energy range of the HPD. A joint detection of such a GRB would therefore, if bright enough, allow to prove or disprove such a model, thereby providing vital insights into the nature of the thermal bump in some GRB spectra.

\begin{figure}
    \centering
    \includegraphics[width=0.45\textwidth]{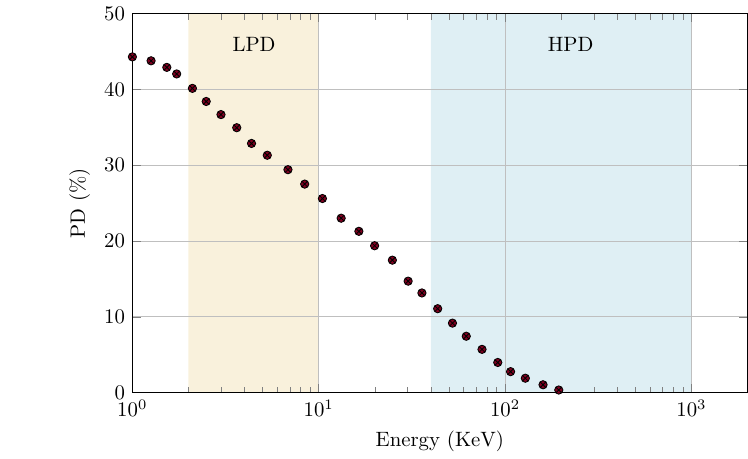}\hspace{3em}
    \includegraphics[width=0.45\textwidth]{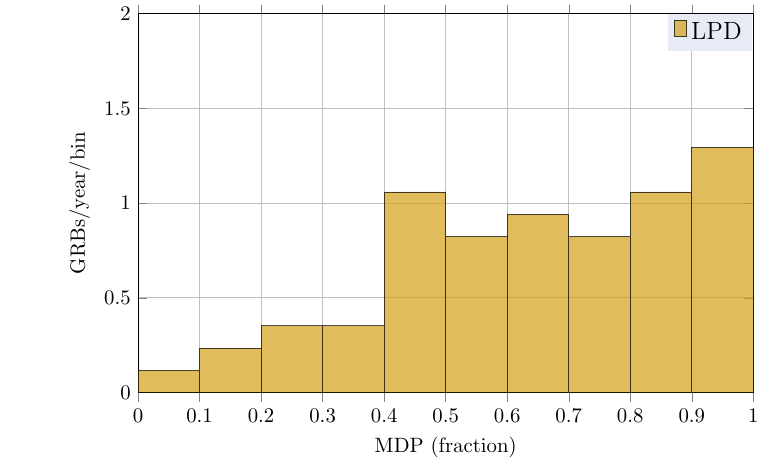}\hspace{3em}
    \caption{\textbf{Left:} The PD as a function of energy predicted by the work in \cite{Lundman_2018} for a dissipative photospheric model. The energy ranges where the HPD and LPD are sensitive to polarimetry are highlighted. \textbf{Right:} The yearly MDP distribution as simulated for the LPD instrument.}
    \label{fig:Lundman_plot}
\end{figure}

To study the possibility of performing constraining energy resolved polarization measurements using both the LPD and the HPD we repeat the simulations using the \textit{Fermi-}GBM catalog but now using both the response of the LPD and the HPD. For this purpose the effective area of the LPD was used, including its dependence on $\theta$ along with the $M_{100}$ of this instrument which is presented in \cite{Yi:2024keb}. The background of this instrument was furthermore studied in detail and details on this can be found in \cite{Feng:2023gdf}. As the two instruments are pointing in the same direction, we assume the incoming angle for both to be equal for each simulated GRB and calculate the MDP for both instruments by folding the spectra through the respective responses. The yearly MDP distribution for the LPD, calculated using 9 years of \textit{Fermi-}GBM data can be seen on the right side of figure \ref{fig:Lundman_plot}. The number of GRBs with an MDP below $100\%$ is about 10 per year, which, given that no GRB polarization measurements exist in this energy range will be of significant importance by itself. In addition, the distribution indicates that, approximately 1 to 2 GRBs will be jointly detected each year with an MDP low energy by both instruments to constrain the emission model shown on the left. For this, the LPD requires an MPD below $40\%$ while the HPD needs one below $10\%$, which occurs for on average 1.4 GRBs per year. With a 2-year lifetime the POLAR-2 mission will therefore provide valuable insights on the validity of this model, while providing the intensive to produce more detailed energy-dependent polarization predictions by other models.

\subsection{BSD Performance}

Preliminary performance studies of the BSD instrument have been conducted using Monte Carlo simulations. The results indicate that the in-orbit background, dominated by the Cosmic X-ray Background (CXB), will induce a total trigger rate of $\sim$5000 counts/s in the 10--100 keV energy band. For GRBs similar to the very weak GRB 170817A, the BSD can achieve a localization precision of approximately $1^\circ$. Furthermore, an observational rate of $\sim$90 GRBs per year is anticipated. The simulated effective area of the BSD imaging detectors as a function of incident photon energy is presented in Fig.~\ref{fig:bsd_arf}.

Apart from simulations, preliminary calibration tests with a BSD prototype detector module have been conducted. These tests demonstrate a detectable energy range from $\sim$10 keV to over 1 MeV, with an energy resolution of $\sim$30\% at 60 keV using a first prototype. Further optimization of the prototype, particularly through optical coupling of the scintillators to the SiPMs is expected to significantly improve this energy resolution.

The design requirement of the BSD for the localization is defined to be $\sim 1^\circ$ for a GRB with a fluence of $10^{-5}$ erg/cm\textsuperscript{2} in the 10--1000~keV range, as this will result in a systematic error on the PD below $1\%$. Based on the first simulations of the BSD, discussed in detail in \cite{BSD_paper} this level of localization is already achieved for GRBs with fluences of $10^{-6}$ erg/cm\textsuperscript{2}. Even for a weak GRB with the brightness of 170817A this remains of the order of $2^\circ$. The spectral requirements for the BSD are more complex to quantify. As the $M_{100}$ depends on the energy, the energy averaged $M_{100}$ used to calculate PD depends on the assumed spectral shape. However, the dependency of $M_{100}$ on energy is strong below 100~keV, while at higher energies $M_{100}$ is relatively  constant with energy. Therefore, the size of the error on the PD from the spectral uncertainty depends strongly on the spectral shape itself. In order to quantify the total dependency, we can look at the typical error on the PD we acquire in the POLAR analysis for simulated GRBs when varying the spectral parameters. We do this using a a sample of 25 simulated GRBs with a fluence of $10^{-5}$ erg/cm\textsuperscript{2} in the 10--1000~keV range. For each we take a random spectral shape from the GBM catalog, add an artificial uncertainty to the spectral parameters and set the PD to $50\%$. First we vary the spectral parameters with the error achievable with POLAR, taken from \cite{Kole2020}, where, when repeating the analysis we find the resulting error on the PD is of the order of $5\%$. Repeating this with the spectral capability of Fermi-GBM (taken again from the catalog) we find the error is below $1\%$. This indicates that the spectral capabilities of the BSD should be equal or better than those of Fermi-GBM. Based on the first simulations, the BSD achieves this through having an effective area typically twice that of GBM and a spectral resolution as a function of energy similar to that from GBM thanks to the use of GAGG scintillators coupled to SiPMs \cite{BSD_paper}.

\begin{figure}[h]
    \centering
    \includegraphics[width=0.7\textwidth]{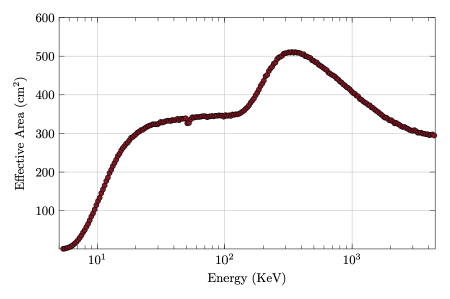}
    \caption{The simulated effective area of the BSD imaging detectors. The effects of the K-edge of Gd around $50.2\,\mathrm{keV}$ can be observed. The dynamic range of the electronics, which is currently being optimized, is not included in the effective area simulation.}
    \label{fig:bsd_arf}
\end{figure}

\section{Conclusions and Discussion}

The POLAR-2 mission includes 3 instruments. The design of each is based on novel technologies and incorporates the lessons learned from previous space missions. In the case of the HPD this includes the use of the successfully tested plastic scintillator array concept, while, based on lessons from the POLAR mission, it is now combined with SiPMs rather than MAPMTs. For this purpose a series of radiation tests of the SiPMs was performed to validate that this relatively novel technology can be used in space for a 2 year mission without the loss of significant radiation damage. This is made possible thanks to an annealing strategy which is of interest for the wider astrophysical community. In addition, the development of the POLAR-2 mission resulted in the a low power, low cost and space qualified electronics system capable of reading out 64 SiPMs with a power of 2 W. This low power allows for the development of a polarimeter with 6400 SiPMs which provides it with an effective area exceeding $2000\,\mathrm{cm}^2$ for spectral events and $1000\,\mathrm{cm}^2$ for polarization events. This not only makes it the most sensitive astrophysical polarimeter ever operated, it also makes it one of the largest $\gamma$-ray detectors to be operational at its time. As such the HPD can, thanks to the fast alert system with which it is foreseen to be equipped, play a vital role in multi-messenger astrophysics by detecting weak GRBs.

A downside of the HPD design is its poor spectral resolution, mostly due to the use of plastic scintillators, and poor localization capabilities. For this purpose it is foreseen to be accompanied by a secondary detector called the BSD which uses the same electronics readout system as the HPD. However, here it is used while coupled to GAGG scintillators. Through this, in combination with a coded mask system, the detector will be capable of producing spectral measurements in the $10\,\mathrm{keV}-1000\,\mathrm{keV}$ energy range and produce localization measurements of 90 GRBs per year with sub-degree precision. This will not only result in significant added value to the real-time transients alerts, but also reduces systematic uncertainties in the polarization measurements of the HPD.

The final instrument on the POLAR-2 mission is the LPD which, for the first time, aims to provide polarization measurements of transient sources in the keV energy range. It does this through a design similar to the detectors employed in the IXPE and PolarLight detectors, but with an optimized gas, a novel ASIC, and a design which makes it capable to be sensitive at larger off-axis angles.

The design of the various instruments is advanced and the mission is currently foreseen to be launched in 2027. As shown in this overview paper, the various components of this payload have been tested both for their performance and qualified to endure both launch and in-orbit conditions for 2 years. The current 2 years of planned operation are set by an agreement between the POLAR-2 collaboration and the organization governing the China Space Station. From a technical perspective, especially when using the discussed SiPM annealing strategies, a longer operation is possible. A potential extension of this time after installation is therefore not excluded. Thanks to the various detailed tests and advances in polarization analysis, it has been possible to produce a relatively detailed predictions on its performance. However, a few vital components of the POLAR-2 mission still require further studies prior to its launch foreseen to be in 2 years. The first of these are optimization of the design for the measurement background and for scattering-induced systematic errors. The background rate used in our estimations is based on extrapolations of that as measured in the POLAR mission. However, the lower energy threshold of POLAR-2 could significantly increase the relative background rate. Further studies of potential passive shielding, as well as trigger strategies are therefore still required. Regarding systematic effects induced by scattering of photons into the polarimeter from passive materials, a tungsten plate is currently foreseen to be used. However, optimization of its thickness to mitigate such effects, as well as additional issues due to, for example, emission lines from the tungsten require further study.

While a real-time alert system of POLAR-2 is foreseen to be used, exact details on its operation, alert criteria and the information it can transmit are ongoing. This is a complex task which requires a detailed understanding of the capabilities of the system, but also simulations of the transient events and the in-orbit conditions. For example, understanding the rate of falsely submitted triggers due to radiation belt induced effects is needed to optimize the system which induces the alert. Furthermore, preliminary processing of the GRB data in the instruments could allow to produce quick alerts to ground containing spectral and localization information, and therefore be valuable. However, this will require detailed tests and optimization of the algorithms to be used.

Finally, while the instruments are designed to produce unprecedented polarization measurements, this can only be achieved after a detailed calibration. An important lesson learned from the last 25 years of gamma-ray polarimetry, is that an understanding of each individual detector element, along with information on its dependency on time and temperature is required. This, for the HPD, requires a pre-flight calibration of the threshold, gain, energy resolution, noise and cross-talk, as well as their dependencies on temperature and radiation damage, of each of the 6400 channels. 

If successfully calibrated and launched, the combination of the 3 payloads will allow to perform GRB measurements within its first 6 months capable of either confirming or disproving $B_{tor}$ synchrotron emission models. Furthermore, by producing the first combined polarization measurements in the X-ray and $\gamma$-ray energy range, the mission will allow for at least 1 energy-resolved polarization measurement per year with an MDP below $40\%$ for both polarimeters. As such it will provide valuable insights on the remaining emission models. Further information on these will be produced through time-resolved polarization measurements, as well as through multi-messenger studies where, for example, the afterglow can be used to constrain jet observation angles. Overall, the mission thereby aims to revolutionize our understanding of GRBs.

		\section*{Author Contributions}
M.K performed the analysis and simulations and was involved in writing and reviewing. N.D.A., J.S., F.X. and H.B.L were involved in writing and reviewing. J.Z. and J.H. were involved in the simulations.

		\section*{Funding}
 F.X. is supported by National Key R\&D Program of China (grant No. 2023YFE0117200), and National Natural Science Foundation of China (grant No. 12373041 and No. 12422306), and Bagui Scholars Program (XF).

		\section*{Acknowledgments}
This research has made use of data obtained through the High Energy Astrophysics Science Archive Research Center (HEASARC), provided by NASA’s Goddard Space Flight Center. 

		\section*{Conflicts of Interest}
 The authors declare no conflict of interest. The funders had no role in the design of the study; in the collection, analyses, or interpretation of data; in the writing of the manuscript; or in the decision to publish the results.

	\small
	\bibliographystyle{scilight}
	%
	\bibliography{example}
	

\end{document}